# Thermalization of Classical Weakly Nonintegrable Many-Body Systems

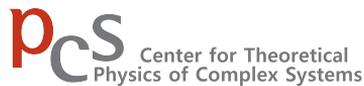

Merab Malishava

Center for Theoretical Physics of Complex Systems, IBS School

University of Science and Technology

A thesis submitted for the degree of

*Doctor of Philosophy*

07 June 2022

# We hereby approve the Ph.D of Merab Malishava

**June 2022**

| | | |
|---|---|---|
| Jung-Wan Ryu | Chairman of Thesis Committee | 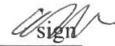 sign |
| Stephano Lepri | Thesis Committee Member | 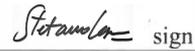 sign |
| Ramaz Khomeriki | Thesis Committee Member | 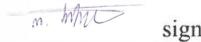 sign |
| Alexei Andreanov | Thesis Committee Member | 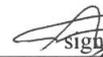 sign |
| Dario Rosa | Thesis Committee Member | 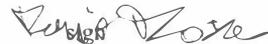 sign |
| Moon Jip Park | Thesis Committee Member | 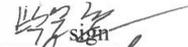 sign |
| Sergej Flach | Thesis Committee Member | 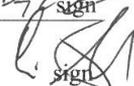 sign |

**UNIVERSITY OF SCIENCE AND TECHNOLOGY**

**Ph.D. Thesis**

# Thermalization of Classical Weakly Nonintegrable Many-Body Systems

**Merab Malishava**

**Basic Science**

**UNIVERSITY OF SCIENCE AND TECHNOLOGY**

**June 2022**

# Thermalization of Classical Weakly Nonintegrable Many-Body Systems

Merab Malishava

A dissertation Submitted in Partial Fulfillment of Requirements for the Degree of Doctor of Philosophy

June 2022

UNIVERSITY OF SCIENCE AND TECHNOLOGY
Major of: Basic Science

Supervisor: Sergej Flach
Co-supervisor: Alexei Andreanov

# ACKNOWLEDGEMENTS

I want to thank my supervisors Alexei Andreanov and Segej Flach for the shared inspiration and guidance on the road to attaining the PhD degree. My thanks goes to the current and former members of PCS for the valuable discussions and insights influencing my professional and personal growth. Special mention goes to our kind and responsive administration and the technical support team. I thank Carlo Danieli and Ihor Vakulchyk for the knowledge they shared and the support during the preparation of the thesis.

# ABSTRACT

We devote our studies to the subject of weakly nonintegrable dynamics of systems with a macroscopic number of degrees of freedom. Our main points of interest are the relations between the timescales of thermalization and the timescales of chaotization; the choice of appropriate observables and the structure of equations coupling them; identifying the classes of weakly nonintegrable dynamics and developing tools to diagnose the properties specific to such classes.

We characterize thermalization of weakly nonintegrable systems using several techniques. We follow a traditional in the field method of computing equilibration times through testing the ergodization hypothesis for a set of chosen observables. We demonstrate how the structure of the network of observables affects the thermalization timescales. We compare these timescales to those of the intrinsic chaotic dynamics given by Lyapunov times. Additionally we follow the evolution of tangent vectors characterizing the direction of propagation of chaos in the system. We quantify the size of tangent vectors and check its scaling against the system size thus characterizing the degree of involvement of the whole system in chaotic dynamics.

Further, we extend our analysis of timescales involved in the thermalization process by computing the whole Lyapunov spectra for large systems. The significant advantage of this technique lies in the independence of Lyapunov spectra from the choice of observables. Additionally the properties of Lyapunov characteristic exponents allow for detection of quasi-conserved quantities emerging as systems are tuned closer to an integrable regime.

Proximity to integrable limit is associated with the rapid growth of thermalization timescales and, thus, potential numerical challenges. We solve these challenges by performing numerical tests using a computationally efficient model - unitary maps. The great advantage of unitary maps for numerical applications is time-discrete error-free evolution. We use these advantages to perform large timescale and system size computations in extreme proximity to the integrable limit. To demonstrate the scope of obtained results we report on the application of the developed framework to Hamiltonian systems.

## 초록

우리는 거시적 수에 해당하는 자유도를 가진 물리계의 약적분불가능역학에 연구를 시행하였다. 주요 관심사는 열화시간척도와 혼동시간척도 사이의 관계와 적절한 관측가능량을 선택하고 그들을 결합하는 방정식의 구조 그리고 약적분불가능역학의 범주를 식별하고 각 범주에 해당하는 특징적인 속성을 진단하기 위한 도구를 개발하는 것이다.

우리는 다양한 기법을 통해 약적분불가능계의 열화를 특성화한다. 주어진 관측가능량 집합에 따라 에르고딕성 가설을 시험하여 평형시간을 계산하는 전통적인 방식을 따른다. 이를 통해 열화시간척도가 관측가능량 네트워크 구조에 따라 어떻게 영향을 받는지 살펴본다. 이러한 시간척도를 혼돈역학계의 고유한 랴푸노프 시간과 비교한다. 더 나아가 우리는 혼돈계의 전파방향을 특정짓는 접선벡터의 진화도 살펴본다. 접선벡터의 크기를 정량화하여 시스템 크기와 함께 비례관계를 살펴보고 전체 시스템의 관여정도를 특정한다.

또한 대형 시스템의 전체 랴푸노프 스펙트럼을 계산하여 열화과정과 관련된 시간척도 분석을 확장한다. 이 기법의 중요한 장점으로 랴푸노프 스펙트럼이 관측가능량의 선택으로부터 독립적이다. 거기에 랴푸노프 특성지수의 성질은 시스템이 적분가능계에 가깝게 조정될 경우에 나타나는 준보존양을 탐지할 수 있다.

적분가능계를 향해 근접하는 것은 열화시간척도의 급속한 증가와 관련이 있으며 이는 곧 수치적으로 큰 도전이나 우리는 효율적인 계산모델인 유니타리맵을 이용하여 해결하였다. 유니타리맵의 큰 장점은 이산시간진화로부터 생성되는 오류로부터 자유롭다. 이러한 장점을 사용하여 적분가능한계에 매우 근접하여 큰 시간규모와 대규모 시스템에서 계산을 수행한다. 얻어진 결과를 입증하기 위해 우리는 해밀턴계에 개발된 프레임워크에 적용하여 보고한다.

# Contents

# List of Figures

# Chapter 1

# Introduction

## 1.1 Motivation

In physics typical development of a theory starts with describing an ideal case scenario. In the case of dynamical systems, one first considers a few-body linear setup before moving towards macroscopic systems with nonlinear laws governing local dynamics. The intuition behind this method is to build upon the existing knowledge so that the understanding of nature develops together with the increasing complexity of observed phenomena. The goal of such an approach is to formulate laws valid for as generic as possible setup and as far as dynamical properties go – to be able to determine the future fate of states observed. The celebrated theories of Newton [1], Einstein [2], Schrödinger [3], and Dirac [4] are the fundamental steps towards achieving the ultimate desire of physicists as a whole – to be able to predict a final state of a system given an initial one.

However, in realistic setups, the knowledge regarding the initial conditions comes with limitations originating from the precision with which the state can be measured. Thus any choice of initial conditions demands a consideration of a set of weakly deviated states. This aspect turns out to be of fundamental importance given the chaotic nature of most physical setups one encounters in practice. States which are initially indistinguishable in terms of a measuring device are separated throughout the evolution with exponentially increasing distance in phase space. Thus after a certain time, almost any final state may have originated from any point in phase space – the memory regarding the initial conditions is lost. In view of this observation, it is reasonable to assume that characterizing chaotic properties of system is of an equal or even greater value than solving for a specific trajectory. The foundations for characterizing the rate of growth of the separation between initially close states were laid out by Lyapunov in 1892 [5] in terms of characteristic exponents and further extended by Kolmogorov and Sinai [6] in terms of Kolmogorov-Sinai entropy.

Further complications come with considering even simple many-body systems. One would assume that the above-mentioned fundamental laws of nature fully describing the behavior of every particle are enough for finding solutions to the many-body problem. These assumptions were tested in the late 19th century by Poincaré who took upon himself a challenge to solve a three body problem of two planets under the gravitational field of the Sun. His work [7] sparked the development of a variety of tools to address chaotic dynamics including notions of integrability and non-integrability, Poincare sections, the famous recurrence theorem [8], etc. However even the three-body problem turned out to be impossible to solve analytically and the resolution came in terms of the Kolmogorov-Arnold-Moser (KAM) theorem [9, 10, 11] which proved the existence of stable quasiperiodic orbits for the case of low planet mass to sun mass ratio.

These complications were addressed to an extent thanks to the tremendous success of statistical mechanics [12, 13, 14, 15]. With the growing number of degrees of freedom, the line of thought diverges from considering single trajectory dynamics in many-body setups. Such systems require statistical description with corresponding macroscopic parameters. The statistical description relies on specific properties of dynamical systems such as ergodicity and mixing [16], which in simplified terms means that a trajectory densely fills the whole phase space. However there exist nonlinear dynamical systems which apparently do not show ergodicity [17]. In fact, the above-mentioned KAM theorem provides a proof for the existence of a non-zero measure set of quasiperiodic orbits in the general case of integrable systems weakly affected by nonintegrable perturba-tion. These orbits do not necessarily visit all points in phase space, resulting in ergodicity breaking dynamics. With increase of the phase space dimensionality the applicability of the KAM theorem quickly becomes limited. Nevertheless questions on the dynamical properties of large systems beyond the limits of the KAM still need to be addressed.

In recent years the development of quantum many-body theory brought up new challenges in understanding the presence and absence of thermalization in certain setups. Specifically many-body quantum systems with impurities demonstrate the absence of thermal properties, which is known as many-body localization. This phenomenon is naturally of interest from the fundamental point of view as many-body interacting systems are expected to show thermalization. The failure to do so opens up potential possibilities to conserve coherent states which is a necessity for sought-after realizations of quantum computers. The classical analog of many-body disordered dynamics also showed the presence of slowing down of thermalization and quickly decreasing conductivity. In view of these findings a deeper analysis and characterization of slow classical dynamics is in demand.

The current thesis is devoted to answering questions posed above. We consider large systems beyond the applicability of KAM regime in extreme proximity to integrable limits where the equilibration dynamics is very slow and the corresponding timescales are diverging. We study these timescales and compare them to those of intrinsic chaotic motion. At the same time we develop tools to characterize and classify such systems thus advancing the knowledge and the currently available methods of analysis in the field.

## 1.2 Overview of weakly nonintegrable dynamics

In this section we provide a brief overview of the equilibration process in weakly nonintegrable systems. In Sec. 1.2.1 We discuss the notion of ergodicity and mixing and elaborate on ergodization and equilibration in physical systems. In Sec. 1.2.2 we define integrability and nonintegrability. We discuss the celebrated KAM theorem and its implications on ergodic dynamics depending on the number of degrees of freedom. Finally in Sec. 1.2.3 we discuss a seminal numerical work on Fermi-Pasta-Ulam-Tsingou (FPUT) model [18] which was the first numerical study of equilibration in a weakly perturbed linear model.

### 1.2.1 Ergodicity, mixing and equilibration

Following the definitions provided in Ref. [19] a state of a physical system with $N$ degrees of freedom is described by specific values attained by a set of independent variables $\{x_n\}_{n=1}^{2N}$, sometimes given in a vector form $\mathbf{x}$. All possible values of $\mathbf{x}$ form a multi-dimensional space $\Omega$ of dimension $2N$ which is referred to as "phase space" [20]. The evolution of such a system is then given as a transformation of a state $\mathbf{x}$ in time. The time $t$ may be continuous $t \in \mathbb{R}$ in which case the transformation is defined as differential equations:

$$\dot{\mathbf{x}} = \mathbf{f}(\mathbf{x}(t)), \tag{1.1}$$

or discrete $t \in \mathbb{Z}$ in which case the transformation is given by a map:

$$\mathbf{x}(t+1) = \mathbf{f}\left(\mathbf{x}(t)\right). \tag{1.2}$$

In practice it is not desirable and frequently impossible to solve for the trajectories $\mathbf{x}(t)$. When describing many-body dynamics one is most of all interested in macroscopic quantities such as pressure, temperature, etc. When making observations using a certain measurement device one rarely performs measurements on a specific state $\mathbf{x_0}(t)$, in fact one does some sort of sampling of a macroscopic function over many microstates [16, 21]. Consider measuring air pressure in a tire – during a measurement process one encounters not single but a macroscopic number of configurations of coordinates and momenta of the constituent air molecules. Thus it is reasonable to assume that the final measured quantity was sampled over all available phase space configurations:

$$O_{\text{measured}} = \int_\Omega O(\mathbf{x}) d\Omega \equiv \langle O \rangle \tag{1.3}$$

From another perspective the time $T$ of measurement procedure is extremely large comparing to microscopic timescales and can be considered infinite. From

this standpoint:

$$O_{\text{measured}} = \lim_{T \to \infty} \frac{1}{T} \int_0^T O(t)dt \equiv \lim_{T \to \infty} \overline{O}_T \qquad (1.4)$$

This intuition is captured by the ergodicity which relates the phase space and time averages. Below we bring a formulation in accordance to [19]:

**Ergodicity:** *The system is ergodic if the phase space average of any observable $O(\mathbf{x})$ is equal to its infinite time average:*

$$\langle O \rangle = \lim_{T \to \infty} \overline{O}_T \qquad (1.5)$$

An equivalent formulation of ergodicity was stated by Maxwell [22]: *"...the system, if left to itself in its actual state of motion, will, sooner or later, pass through every phase which is consistent with the equation of energy."*

Note, that our definitions of averages relate to continuous time and phase space. In time-discrete cases, the integrals are trivially replaced by sums. We also assume that the total volume of the phase space equals unity.

Rigorous definitions and proofs of ergodicity and consequent aspects were laid out by Birkhoff [23] and von Neumann [24]. Strictly speaking, Hamiltonian systems are not ergodic due to the conservation of energy. However, the intuitive ergodic hypothesis for dynamics of isolated Hamiltonian systems on hypersurfaces of constant energies was initially introduced by Boltzmann [13] and extensively utilized in the formulation of statistical mechanics. We refer to a process of time averages approaching the phase space averages as the **ergodization**.

The assumption of ergodicity is of great importance for computing observables at thermodynamic equilibrium:

**Thermodynamic equilibrium:** *The system is at thermodynamic equilibrium if its macroscopic observables do not change in time. The values of macroscopic observables in thermodynamic equilibrium equal to their phase space average* [21].

From ergodic hypothesis Eq. (1.5) it follows that one is able to compute time averages and extract the timescale at which the condition (1.5) is satisfied. One may call this the equilibration time. Another variant frequently employed for extracting equilibration time stems from the **equipartition theorem** which states that the energy per degree of freedom averaged over time of freedom equals the energy density in the system.

A stronger than ergodicity property frequently attributed to the phase space dynamics of physical systems is that of *mixing:*

**Mixing:** *The system is said to be mixing if almost any two trajectories in phase space come arbitrarily close during the evolution.*

The difference between ergodicity and mixing is subtle but important. Mixing necessarily implies the decay of correlations between any two states over time. Ergodicity does not. Consider two trajectories separated in phase space by some distance $d$ at initial time of the evolution $t = 0$. During the evolution both trajectories may cover the whole phase space; that would correspond to ergodicity. However the distance $d$ might stay constant during this process. In such case the dynamics would not be mixing. The correlations between two initial states would not decay in time. For the detailed discussion we refer readers to the Ref. [19].

The mixing dynamics is necessarily ergodic [19, 25, 26, 27]. However, the converse is not true which is demonstrated by a simple example of uniform rotation of a circle – such a transformation is ergodic, but not mixing [19]. In regards to physical models, an example of ergodic but not mixing dynamics is a set of uncoupled harmonic oscillators with incommensurate frequencies. The energy of each oscillator is conserved, which restricts the available phase space. However, in that phase space, the trajectories will fulfill the ergodic criteria of passing arbitrarily close to any point and show seeming equilibration [28]. This linear setup, however, is far from realistic physical considerations where nonlinear interaction laws govern the dynamics and make it mixing. A reasonable approach to study equilibration in realistic setups would be to start with linear, idealized dynamics and introduce a nonlinear, chaos-inducing perturbation. In the subsequent sections, we will consider a theory 1.2.2 and a numerical experiment 1.2.3 which laid the foundation for understanding the equilibration of weakly perturbed dynamical systems.

### 1.2.2 Kolmogorov-Arnold-Moser theory

The celebrated KAM theorem deals with the notions of integrable and nonintegrable Hamiltonian systems. Hamiltonian dynamics of an isolated system with $N$ degrees of freedom is described by a set of $2N$ time dependent variables $\{q_i(t),\ p_i(t)\}$ which are evolved in $2N$ dimensional phase space according to the Hamilton equations of motion conserving the value of the energy function $H(\boldsymbol{q},\boldsymbol{p})$:

$$\dot{\boldsymbol{q}} = \frac{\partial H}{\partial \boldsymbol{p}}, \quad \dot{\boldsymbol{p}} = -\frac{\partial H}{\partial \boldsymbol{q}}. \tag{1.6}$$

**Integrability:** *The system is coined integrable if there exists a transformation $\{q_i, p_i\} \to \{\varphi_i\, I_i\}$ such that new variables obey the following equations of motion:*

$$\dot{\boldsymbol{I}} = -\frac{\partial H}{\partial \boldsymbol{\varphi}} = 0, \quad \dot{\boldsymbol{\varphi}} = \frac{\partial H}{\partial \boldsymbol{I}} = \boldsymbol{\omega}(\boldsymbol{I}). \tag{1.7}$$

The new variables are referred to as *action-angle variables*. The trajectories of integrable systems wind up on surfaces of multidimensional tori parametrized by actions $\boldsymbol{I}$ with constant angular velocities $\boldsymbol{\omega}(\boldsymbol{I})$. If for any frequencies $\omega_i = a\omega_j$ with proportionality coefficient $a \in \mathbb{Q}$ rational, then the dynamics is *periodic*. In case of rationally independent frequencies $\omega_i$ the motion is called *quasi-periodic*. The quasiperiodic trajectories fill the multidimensional tori densely and in that sense may even be considered pseudo-ergodic. This is exactly the case in the model of uncoupled harmonic oscillators with incommensurate frequencies discussed above. However, for all intents and purposes we consider systems with a macroscopic number of integrals of motion as nonergodic.

For most physical systems a Hamiltonian may be written as follows:

$$H = H_{\text{int}}(\boldsymbol{I}) + \varepsilon H_{\text{nonint}}(\boldsymbol{\varphi}, \boldsymbol{I}). \tag{1.8}$$

Here $H_{\text{int}}$ represents an integrable part of the Hamiltonian (for instance kinetic energy part of interacting gas molecules), $H_{\text{nonint}}$ is a nonintegrable part (interaction) and $\varepsilon$ is the perturbation parameter. One would expect that a system governed by such Hamiltonian would explore the available phase space as its dynamics is no longer confined to multidimensional tori. The Kolmogorov-Arnold-Moser theorem addresses this intuitive expectation:

**KAM theorem:** *For a nonintegrable Hamiltonian system $H = H_{int} + \varepsilon H_{nonint}$ there exists sufficiently small value of parameter $\varepsilon$ such that a finite measure set of quasiperiodic trajectories of initially unperturbed integrable Hamiltonian $H_{int}$ stays quasiperiodic.*

As stated above, historically the KAM theorem [9, 10, 11] was intimately related with the three-body problem of a massive star and two planets. Poincare showed that such a problem may be simplified to a Hamiltonian of the type given by Eq. (1.8) where the perturbation is given by the ratio of the mass of planets to that of a star [29].

From the first glance, the KAM theorem invalidates the ergodic hypothesis for sufficiently small perturbations to integrable dynamics. However its consequences are not taken into account in the statistical mechanical description of physical systems. Why? The reason lies in the limited applicability of the KAM theorem. In particular, it has been shown that in order for the results of the theorem to hold the strength of a nonintegrable perturbation $\varepsilon$ has to rapidly decay with the number of degrees of freedom [30, 31, 32]. For lattice systems the scaling of the critical value of the nonintegrable perturbation with the system size was shown in Ref. [31]:

$$\varepsilon \sim N^{-b} \text{ , for some large number b.} \quad (1.9)$$

Thus for the purposes of statistical description of the ergodic phase space dynamics in weakly nonintegrable systems with a large number of degrees of freedom the KAM theorem does not seem to pose a threat. In its yearly days, however, the KAM theory has been suggested as the explanation for phenomena exhibiting ergodicity breaking behavior. A notable example came in terms of the first numerical experiment aiming to demonstrate energy equipartition in a physical system.

### 1.2.3 Fermi-Pasta-Ulam-Tsingou Model

In the 1950s after the completion of MANIAC-I computer the scientific community was ready to perform numerical experiments to test some basic assumptions of statistical mechanics such as equipartition of energy, ergodicity, etc. In the pioneering work by Fermi, Pasta, Ulam and Tsingou the following Hamiltonian was considered [18, 33, 34]:

$$H = \sum_{n=1}^{N} \left[ \frac{1}{2} p_n^2 + \frac{1}{2}(q_{n+1} - q_n)^2 + \frac{\alpha}{3}(q_{n+1} - q_n)^3 \right]. \tag{1.10}$$

The boundary conditions are fixed $q_0 \equiv q_{N+1} = 0$. The Hamiltonian (1.10) describes a one dimensional chain of unit masses with nearest neighbors connected by identical nonlinear springs. The strength of nonlinearity is controlled by a parameter $\alpha$.

The essential idea of the experiment is to prescribe some small values to nonlinear terms and observe energy equipartition between the unperturbed harmonic modes given as:

$$Q_k = \sqrt{2/N} \sum_{n}^{N} q_n \sin[kn\pi/(N+1)]$$
$$P_k = \sqrt{2/N} \sum_{n}^{N} p_n \sin[kn\pi/(N+1)]. \tag{1.11}$$

In terms of the normal modes the Hamiltonian above can be rewritten as a set of harmonic oscillators with frequencies $\omega_k = 2\sin[k\pi/2(N+1)]$ weakly perturbed by an anharmonic term:

$$H = \frac{1}{2} \sum_{k=1}^{N} \left[ P_k^2 + \omega_k^2 Q_k^2 \right] + \alpha \sum_{k,k_1,k_2=1}^{N} I_{k,k_1,k_2} Q_k Q_{k_1} Q_{k_2}. \tag{1.12}$$

The constants $I_{k,k_1,k_1}$ are a result of overlap of normal modes and are sometimes referred to as "overlap integrals". The typical form of such constants is given below in Sec. 1.4.2.

The energy of each mode is given as $E_k = P_k^2 + \omega_k^2 Q_k^2$ and due to weak nonlinear coupling the total energy $E_{\text{tot}} \approx \sum_k E_k$. The FPUT fixed the system size $N = 32$, and the nonlinearity parameter $\alpha = 1/4$. The initial conditions with the first mode excited $q_n \sim \sin[\pi n/(N+1)]$. With the energy density of the order of $E_{tot}/N \approx 10^{-3}$ the nonlinear coupling in Eq. (1.12) plays the role of weak perturbation to linear dynamics [33]. As per the ergodic hypothesis the expected result was the distribution of energy through the modes and for the trajectory to visit the whole available phase space. The result was an unexpected recurrence of the trajectory back to initial conditions after exciting just few neighboring modes, see Fig. 1.1. The energy of initially excited mode recovered back within a window of $3\%$ [33]. Moreover, the energy stays localized within the first four modes of the linear system with other modes staying at rest for the whole time of the evolution [18].

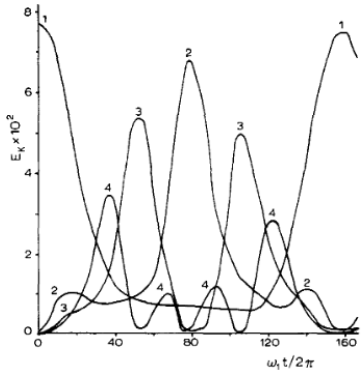

Figure 1.1: The demonstration of the FPUT recurrence taken from [33]. Curves 1 to 4 represent the evolution in time of the energies of the first 4 normal modes.

The surprising result challenged the assumptions regarding the ergodicity in nonlinear interacting systems and attracted a lot of attention from the statistical mechanics community at the time and a number of hypotheses to explain the behavior emerged [33]. A simple explanation of the results by FPUT being an evidence for Poincare recurrence were dissolved almost immediately after the experiment as the timescale of the recurrences scale with system size $N$ as a power law, while Poincare times scale exponentially [35]. The doubts regarding thermalization of one-dimensional systems were also disproved [36]. The most believed hypothesis that the integration time by FPUT was simply not enough to witness the thermalization was also tested. These tests not only failed to observe thermalization but found the so called "super periods" of recurrences on larger timescales [37].

After aforementioned intuitively simple explanations of the phenomenon were addressed there emerged two notable directions to shed light on the FPUT paradox. The first, developed by Zabusky and Kruskal considers the limit of a continuous system which leads to mapping to Korweveg-De Vries (KdV) equation. The KdV equation assumes localized in space soliton solutions [38]. For the large wave length cases the continuous system approximation could be a reasonable way to explain recurrences, however in the opposite scenario this approach is not applicable. Thus the mapping onto the KdV equation does not constitute a general explanation, but merely an approximation. Another notable approach by Chirikov and Izrailev suggested a nonlinearity strength threshold [39]. For large enough nonlinearity or energy density the FPUT recurrence is destroyed and expected thermal behavior emerges [40]. This particular approach bases itself in the realm of the KAM theorem discussed above. Essentially the claim is that the parameters chosen in the FPUT experiment result in a small enough perturbation to the integrable Hamiltonian for the KAM theorem to be applicable. However, some works have shown the dynamics possessing chaotic properties but still staying localized among the four lowest number modes [41, 42], which seemingly contradicts KAM. Finally the FPUT model has been shown to admit exact stable breather solutions localized in the k-space [43, 44]. The FPUT initial conditions may be viewed as a deviation from such an exact solution with extremely large thermalization times.

The origins of the FPUT paradox are still discussed. This is amplified by the fact that it has been nearly 70 years since the original report by Fermi, Pasta, Ulam and Tsingou but the model has yet to show thermalization for the originally chosen parameters and initial conditions. The recent numeric estimation of possible equilibration time is $T_E = 10^{14}$ [45] – a truly astronomic time. It is somewhat ironic that the model chosen to be put on display as the triumph of the statistical mechanical description of equilibration turned out to be exactly the opposite.

## 1.3 Timescales and methods of their computation

The study of equilibration of any physical system is first and foremost a study on the timescales. Typically to constitute equilibration the ergodicity time with condition given in Eq. (1.5) or equipartition time is measured. In the FPUT experiment the expected (although not observed) equilibration time is associated with the equipartition of the energy through the normal modes of the linear harmonic oscillator chain. On the other hand there exist timescales associated with chaotic dynamics. These are given by Lyapunov times and essentially relate to the rate of mixing in phase space. Most of the time the shortest time of chaotization, i.e. the smallest Lyapunov time is measured as a baseline for chaoticity related dynamics.

The relation between the two timescales is a frequent point of interest in the studies of equilibration in dynamical systems and consequently so are the methods for their computation.

Below we cover some standard methods for the computation of the relevant timescales. Section 1.3.1 covers the ergodicity related timescales. There is no standard or even a dominant technique for extracting the ergodization time. The condition given in Eq. (1.5) deals with infinite time averages which are practically inaccessible. Thus some workaround techniques involving cutoffs are introduced. The computation of Lyapunov time also relies on theory involving infinite time computations. However there exist standard procedures to assess it using finite time computations. We cover the aspects of tangent vector dynamics as well as the definitions for Lyapunov exponents in Sec. 1.3.2.

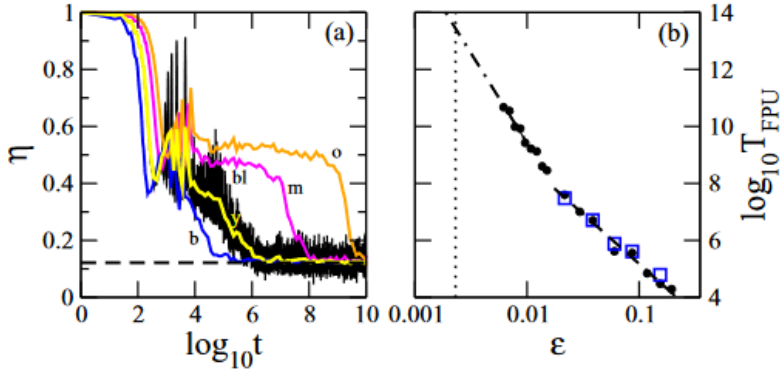

Figure 1.2: An example of equilibration time computation in the FPUT model taken from Ref. [45]. (a) Evolution of the window averaged $\eta(t)$ for different values of energy density (colored). Instantaneous evolution of $\eta(t)$ for one value of energy is given in (black). The dashed line indicates the threshold value $\eta_c$. (b) Evolution times $T_E$ extracted when $\eta(t) = \eta_c$ against the energy density $\varepsilon$.

### 1.3.1 Equilibration time

To assess equilibration time in Hamiltonian systems one frequently tests equipartition theorem for the set of energies $\{E_n(t)\}$ with the assumption $\sum_n E_n(t) = const$. Typical examples of such set of energies are local kinetic or potential energies, energies of normal modes, etc. One widespread measure of equipartition initially introduced in Ref. [46] is given in terms of the spectral entropy:

$$S(t) = -\sum_n^N \epsilon_n(t) \log \epsilon_n(t). \qquad (1.13)$$

Here $\epsilon_n(t) = E_n(t)/\sum_n E_n(t)$ are the renormalized energies. The spectral entropy is zero when only one element of the set excited, and reaches its maximal value $S_{max} = \log N$ in case of complete equipartition of the total energy among all degrees of freedom. To avoid the dependence on the number of degrees of freedom, the normalized quantity is introduced:

$$\eta(t) = \frac{S(t) - S_{max}}{S(0) - S_{max}}. \qquad (1.14)$$

The function $\eta(t)$ is bounded between zero (perfect equipartition), and one (complete "localization" in one degree of freedom). At this point the cutoff value $\eta_c$ is introduced. In some cases the phase space average $\langle\eta\rangle$ can be computed analytically and chosen as the cutoff value [45]. In others this value is chosen ad hoc [47, 48]. The first time $T_E$ during the evolution when $\eta(T_E) = \eta_c$ is taken as the equilibration time, see Fig. 1.2 for an example.

An alternative method for extracting the equilibration time pertains the analysis of distributions of finite time averages (FTAs). This method revolves around testing the ergodic hypothesis and has recently been used in multiple works on characterization of weakly nonintegrable dynamics [49, 50, 51]. The ergodic hypothesis deals with the infinite time averages which are numerically inaccessible. By performing the analysis of statistical properties of FTA distributions one is able to extract the timescales of thermalization as well as find the rates at which the time averages converge to the phase space averages. First, for a selected observable $O(t)$ a finite time average is defined:

$$\overline{O}_T = \frac{1}{T}\int_{t=0}^{T} O(t)dt. \quad (1.15)$$

The next step is to construct a set of FTAs. For this a large number $M$ of trajectories is evolved and correspondingly finite time averages are measured in each case. As a result a set of FTAs is obtained $\{\overline{O}_T\}_M$ with respective probability density function $\varrho(\overline{O}_T)$. For each value of the integration time $T$ the distribution is typically characterized by its first and second moments $\mu_1(T)$ and $\mu_2(T)$. Since averaging over time commutes with averaging over trajectories, the first moment is time independent and coincides with the phase space average:

$$\mu_1(T) = \frac{1}{M}\sum_i^M \left(\frac{1}{T}\int_t^T O_i(t)dt\right) = \frac{1}{T}\int_t^T dt\left(\frac{1}{M}\sum_i^M O_i(t)\right) = \langle O\rangle \quad (1.16)$$

The last equality stems from the fact that at any given moment in time the instantaneous values of $M$ observables $\{O(t)\}_M$ represent nothing but a sample from the phase space with a corresponding average value $\langle O\rangle = \frac{1}{M}\sum_i O_i(t)$. As an example, if a set of energies is chosen as observables, the phase space average can be computed from the Gibbs distribution. The second moment $\mu_2(T)$ is still time dependent, since there is no permutation of integrals such as in the Eq. (1.16).

As per the ergodic hypothesis for each FTA we expect $\bar{O}_{T\to\infty} = \langle O \rangle$. The distribution function $\varrho(T)$ is therefore expected to peak around the phase space average, reducing its variance to zero and thus approaching a delta-function for infinite averaging times:

$$\varrho(\bar{O}_{T\to\infty}) = \delta(\bar{O}_{T\to\infty} - \langle O \rangle). \tag{1.17}$$

To study the convergence one computes the second moment $\mu_2(T)$, variance $\sigma^2(T)$ or a related quantity such a scale free fluctuation index $q(T) = \sigma^2(T)/\mu_1^2$ of the distribution $\varrho(\bar{O}_T)$. These quantities essentially aim at describing the width of the distribution and are expected to tend to zero as the averaging time $T$ is increased. Naturally the condition given in Eq. (1.17) will never be satisfied exactly due to finite time computations. This again leads to implementation of certain cutoff values for the width of the distribution $\varrho$ to extract the equilibration time $T_E$ [50]. Specific applications of this particular method is presented in following sections.

### 1.3.2 Lyapunov time

From the standpoint of the evolution in phase space the chaotic dynamics is described by its own characteristic timescale. Loosely speaking, given two trajectories $\mathbf{x_1}(t)$ and $\mathbf{x_2}(t)$ that are close at $t = 0$ the distance between them $d(t) = ||\mathbf{x_2}(t) - \mathbf{x_1}(t)||$ will grow exponentially fast. The rate of growth is given by a characteristic exponent $\Lambda$ and a corresponding time $T_\Lambda = 1/\Lambda$. The characteristic exponents were originally introduced by Aleksandr Lyapunov in his studies of the stability of nonstationary solutions of ordinary differential equations [5], and are named after him. The largest Lyapunov exponent is used to detect chaotic dynamics and the shortest timescale of exponential divergence of trajectories. Naturally this timescale is compared with the ergodization time discussed above. Below we cover the basis for its computation with further expanding this approach to computation of the entire Lyapunov spectrum in the Sec. 2.3.1.1.

For a reference trajectory $\mathbf{x}(t)$ one may define tangent vectors (sometimes called variational or deviation vectors):

$$\mathbf{w}(t) = \delta\mathbf{x}(t), \tag{1.18}$$

and from equations of motion the Jacobi matrix:

$$\frac{\partial \mathbf{f}}{\partial \mathbf{x}} = \begin{pmatrix} \frac{\partial f_1}{\partial x_1} & \frac{\partial f_1}{\partial x_2} & \cdot & \cdot & \cdot & \frac{\partial f_1}{\partial x_{2N}} \\ \cdot & \cdot & & & & \cdot \\ \cdot & \cdot & & & & \cdot \\ \cdot & \cdot & & & & \cdot \\ \frac{\partial f_{2N}}{\partial x_1} & \frac{\partial f_{2N}}{\partial x_2} & \cdot & \cdot & \cdot & \frac{\partial f_{2N}}{\partial x_{2N}} \end{pmatrix} \tag{1.19}$$

The tangent vectors evolve in time according to the Jacobi matrix. In time continuous case:

$$\dot{\mathbf{w}} = \frac{\partial \mathbf{f}}{\partial \mathbf{x}} \mathbf{w}(t), \tag{1.20}$$

or in time discrete case:

$$\mathbf{w}(t+1) = \frac{\partial \mathbf{f}}{\partial \mathbf{x}} \mathbf{w}(t), \tag{1.21}$$

Note that the equations of motion above are linear in $\mathbf{w}(t)$. The coefficients are time dependent since the Jacobi matrix (1.19) is defined for a reference trajectory $\mathbf{x}(t)$ at every time $t$ of the evolution. For this reason integrating the tangential equations of motion requires the knowledge of a trajectory $\mathbf{x}(t)$. In practice this means evolving $\mathbf{x}(t)$ and $\mathbf{w}(t)$ simultaneously. For instructive examples on the integration of tangent equations in simple systems such as the Henon-Heiles [17] model we refer readers to [52].

The absolute value of a tangent vector $||\mathbf{w}(t)||$ represents the distance between a reference trajectory $\mathbf{x}(t)$ and a trajectory which was originally infinitesimally close to it. The coefficient $r(t) = ||\mathbf{w}(t)||/||\mathbf{w}(0)||$ is called the coefficient of expansion in the direction of $\mathbf{w}$. The following limit value is called the **largest Lyapunov characteristic exponent** or **1–LCE**:

$$\lim_{t \to \infty} \frac{1}{t} \log r(t) \equiv \Lambda. \tag{1.22}$$

A deeper investigation of tangent dynamics reveals not one, but a spectrum of Lyapunov exponents. The details on the Lyapunov spectrum are covered in Sec.

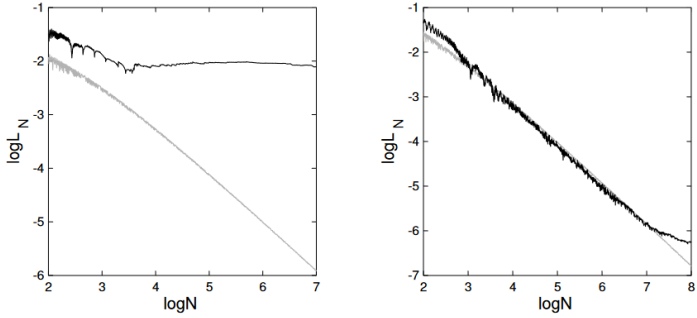

Figure 1.3: An example of computation of the transient Lyapunov exponent $X(t)$ (denoted as $L_N$, where $N$ is equivalent of time $t$) in a 4-dimensional map [17]. The gray curves show the case of the regular orbit and a corresponding decay towards $X(t \to \infty) = 0$. The black curves show the chaotic motion with eventual saturation on smaller times on the left and larger times on the right. Typically the last value of $X(t_{max})$ is taken as the value of the Lyapunov exponent. The figure is taken from [53].

2.3.1.1. Thus far we point out that the exponent defined above is the largest in the spectrum and frequently written as $\Lambda_{max}$. In chaotic systems the value $r(t)$ grows exponentially in time $r \sim e^{\lambda_t t}$. The exponent $\Lambda$ is nothing but an infinite time limit of momentary growth exponent $\lambda_t$. The time $T_\Lambda = 1/\Lambda$ is called the **Lyapunov time**. Essentially it is a timescale corresponding to the deviation of the initially close trajectories.

The computation of the Lyapunov exponent requires integrating a trajectory $\mathbf{x}(t)$ according to Eq. (1.1) together with a tangent vector $\mathbf{w}(t)$ according to Eq. (1.20) with subsequent extraction of the coefficient $r(t)$. By definition the Lyapunov exponent is obtained in the infinite time limit. Thus the so called transient Lyapunov exponent $X(t) \equiv \frac{1}{t} \log r(t)$ is computed. The utility of the 1–LCE is the identification of chaotic dynamics from $\Lambda_{max} > 0$ criterion. In regular dynamics case there is no exponential deviation of trajectories. Thus the Lyapunov exponent is vanishing $\Lambda_{max} = 0$. In practice this means the saturation (in chaotic case) or decay (in regular case) of the transient exponent $X(t)$. In Fig. 1.3 we show typical results of computation of the largest LCE for chaotic and regular setups. The algorithms used for the computation of the Lyapunov time can be found in Refs. [52, 54].

There exist cases of ergodic dynamics without chaoticity [55]. Evaluating the largest LCE allows for differentiation of such cases from truly chaotic motion with mixing. Quasiperiodic dynamics densely winding on tori in phase space are ergodic, however the 1–LCE will be vanishing indicating the hidden integrability which obstructs thermalization. Such findings cannot be made by analyzing observable dynamics.

## 1.4 Recent developments

As was well demonstrated by examples of the KAM theory and the FPUT experiment, a failure to observe ergodicity frequently attracts great attention from statistical mechanics and dynamical systems community. The recent works on the absence of thermalization in quantum systems, known as many-body localization [56, 57, 58] amplified the ongoing discussions. In particular, some connections between quantum and classical nonthermal behavior were made in Refs. [59, 60]. This was supplemented by the reports on slow glassy dynamics which are characterized by the anomalously slow decay of correlations [61, 62] and thermal conductivity [63]. The studies on slowing down of ergodization resulted in the development of a framework of methods for probing and classification of weakly nonintegrable dynamics. Below we will elaborate on this framework relying on Refs. [49, 50, 51, 62].

### 1.4.1 Dynamical glass

The term 'dynamical glass' refers to the dynamics with large characteristic equilibration timescales. In particular such regime was reported in lattice setups that turn integrable in the limit of vanishing couplings [49, 62]. As opposed to the FPUT setup discussed above where the nonlinear interaction serves as a perturbation to the integrable dynamics, here the perturbation comes in terms of weak

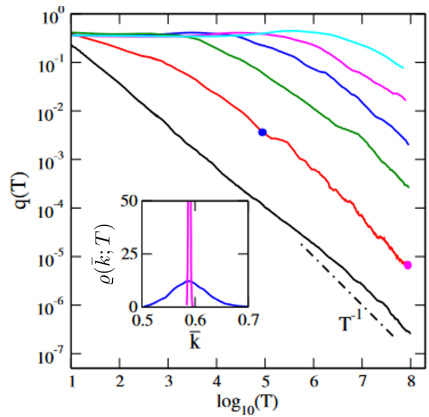

Figure 1.4: The dependence of the fluctuation index $q(T)$ corresponding to the set of finite time averages $\{k_n\}$ as the system is tuned closer to the integrable limit. The energy density varies from $h = 0.1$ (black) to $h = 8.4$ (cyan). The inset shows the peaking over time of the distribution $\varrho(\bar{k}, T)$ around average value of the kinetic energy. The data is measured at two points of the red curve corresponding to $h = 1.2$. The figure is taken from Ref. [49].

coupling of local actions. An instructive case is given by the Josephson junctions networks:

$$H = \sum_{n=1}^{N} \left[ \frac{1}{2} p_n^2 + E_J \left(1 - \cos\left(q_{i+1} - q_i\right)\right) \right]. \tag{1.23}$$

The model represents the dynamics of a chain of $N$ superconducting islands with weak nearest neighbor coupling $E_J$ in its classical limit. The Josephson junction model is equivalent to a coupled rotor chain. The equations of motion for the set of canonical variables $\{q_n, p_n\}$ are as follows:

$$\dot{q}_n = p_n, \qquad \dot{p}_n = E_J \left[ \sin\left(q_{n+1} - q_n\right) + \sin\left(q_{n-1} - q_n\right) \right]. \tag{1.24}$$

If the model is tuned to a regime of vanishing coupling $E_J \to 0$ or alternatively large energy density $h = H/N \to \infty$ the momenta $p_n$ turn to actions of the integrable limit and each element of the set of local kinetic energies $k_n = p_n^2/2$ is conserved. In this limit the system turns into a set of uncoupled superconducting grains described by the integrable Hamiltonian $H_0 = \sum_n \frac{1}{2} p_n^2$. Introducing weak coupling $E_J$ (or alternatively increasing energy density $h$) results in a nonintegrable perturbation to the Hamiltonian $H_0$ thus coupling the actions. It is possible to examine the dependence of ergodicity timescales on the perturbation parameter $E_J$. In Ref. [50] the authors follow the approach presented in Sec. 1.3.1 and study the ergodization of local kinetic energies $\{k_n\}$ using the width of the distribution of finite time averages $\varrho(\overline{k}; T)$, where $T$ is the integration time and $\overline{k}$ is a set of finite time averages $\overline{k}_n(T) = (1/T) \int_0^T k_n(t) dt$. To characterize the width of the distribution the scaleless quantity is used $q(T) = \sigma^2(T)/\mu_1^2$ where $\mu_1$ is the first moment of the distribution $\varrho(\overline{k}, T)$. In Fig. 1.4 we show the behavior of the $q(T)$ as the energy density $h$ varies. Ergodicity requires $q(T)$ to decay in time as the distribution $\varrho(\overline{k}, T)$ peaks around the average value (see the inset). On the other hand in the limit $h \to \infty$ the fluctuation index $q(T)$ is expected to show constant behavior. For the values of energy density between these two regimes there exists a flat behavior (signature of weak nonintegrability) followed up by eventual decay due to ergodicity.

Next step is to extract the ergodization timescale $T_E$, that is the timescale on which the decay starts showing, and compare it with the chaoticity timescales given by the Lyapunov time $T_\Lambda$. The extraction of the ergodization timescales $T_E$ involves the rescaling of the curves found in Fig. 1.4 (see Ref. [49] for details). The computation of Lyapunov time was done according to the procedure presented in Sec. 1.3.2. The discovery lies in the surprising result $T_E \gg T_\Lambda$ as the system is tuned towards the integrable limit. For the set of parameters used in the Ref. [49] the ratio of timescales reaches up to $T_E/T_\Lambda \approx 10^6$. This is the dynamical glass. This result signifies the presence of other timescales relevant for the thermalization.

To shed light on the microscopic origins of the dynamical glass the analysis has been extended to the so called excursion times, i.e. the time events $\Delta t^\pm \equiv \tau^\pm$ that an observable takes the value above $k_n(\tau^+) > \langle k \rangle$ or below $k_n(\tau^-) < \langle k \rangle$ the phase space average. Throughout the evolution an observable fluctuates around it's average and thus the distribution of excursion times $P_\pm(\tau)$ may be collected. It turns out that in proximity to integrable limit the number of long lasting exci-tations grows quickly. In particular the tails of the distribution show polynomial fat tail $P_\pm \sim \tau^{-2}$ behavior with diverging mean $\mu_\tau$ and standard deviation $\sigma_\tau$. In Ref. [49] the relation between ergodization times and the excursion times has been found: $T_E \sim \sigma_\tau^2/\mu_\tau$. The Fig. 1.5(a) shows fluctuations with large excursion times lasting up to $\tau = 10^5$. Another way to describe local dynamics is through chaotic resonances. The momenta $p_n$ are equivalent to frequencies $\omega_n$ of action-angle representation. The two neighbors are defined to be at resonance if the frequency difference $\Delta_n = |\omega_{n+1} - \omega_n|$ for two neighboring sites is less than one: $\Delta_n < 1$ and $\Delta_{n+1} < 1$ [59, 64, 65]. In Fig. 1.5(b) the spots with chaotic resonances are shown with black. We notice the rare resonances, as well as some regions which are nearly frozen for $\tau \approx 10^5$. Such nearly frozen dynamics is characteristic of dynamical glass regime [45, 62]. The link between fast growth of ergodization timescale $T_E$ and increasingly rare chaotic resonances has been suggested, however the strict proof is yet to be shown [49].

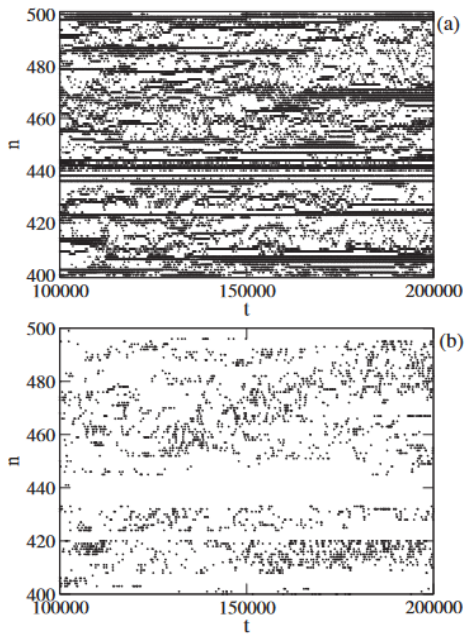

Figure 1.5: Longlasting excitations in Josephson junction chains [49]. (a) Spatiotemporal evolution of a part of the chain. Black corresponds to $k_n > \langle k \rangle$, white corresponds to $k_n < \langle k \rangle$. (b) Evolution of nonlinear resonances. Black corresponds to the presence of resonances. For both cases $E_J = 1$, $h = 5.4$.

### 1.4.2 Networks of observables

As discussed, the dynamical glass regime is relevant in the context of lattice systems with vanishing inter-site couplings, such that the local dynamics is quasi-conserved upon approaching the integrable limit. Such case differs intuitively from the one observed in FPUT or alike models where the perturbation to integrable limit comes in terms of the nonlinearity coupling normal modes of a linear system. This difference was highlighted in Ref. [50] that extended the approach discussed above to the case of weakly nonlinear systems and the two distinct weakly nonintegrable regimes were separated in two different classes according to the displayed equilibration dynamics. These classes encapsulate a great number of physical models. To demonstrate this approach we follow Ref. [50] and use the example of the Klein-Gordon model:

$$H_{\mathrm{KG}} = \sum_n \left[ \frac{1}{2} p_n^2 + \frac{\varepsilon}{2}(q_{n+1} - q_n)^2 + \frac{1}{2} q_n^2 + \frac{1}{4} q_n^4 \right]. \quad (1.25)$$

Here the coupling strength $\varepsilon$ and the energy density $h$ are the parameters relevant for the future discussion. The Hamiltonian $H_{\mathrm{KG}}$ possesses two qualitatively distinct integrable limits. For $h = const$, $\varepsilon \to 0$ the system turns to the chain of uncoupled anharmonic oscillators with the energy of each oscillator turning to an integral of motion. The coupling potential in this case acts as the perturbation to an otherwise integrable system. The equations of motion for the coordinates $q_n$ are as follows:

$$\ddot{q}_n = -q_n - q_n^3 + \varepsilon(q_{n+1} + q_{n-1} - 2q_n) \quad (1.26)$$

In the weakly nonintegrable regime $h = const$, $\varepsilon \ll 1$ the exchange of energy between oscillators happens locally. This regime is analogous to the one discussed in the previous section 1.4.1. Here the perturbation to integrable dynamics is local. Each oscillator is coupled to a finite number of neighbors.

The qualitatively distinct is the limit $h \to 0$, $\varepsilon = const$. In this case the nonlinear term in the Hamiltonian (1.25) vanishes. Similarly to the FPUT model the energy of the normal modes $\{Q_k, P_k\}$ (1.11) are the integrals of motion with the following energies:

$$E_k = \frac{P_k^2 + \Omega_k Q_k^2}{2} \quad , \quad \Omega_k = \sqrt{1 + \varepsilon \omega_k},$$
$$\omega_k = 2\sin\frac{\pi k}{2(N+1)}. \tag{1.27}$$

The normal modes are coupled according to the equations below:

$$\ddot{Q}_k + \Omega_k^2 Q_k = -\frac{1}{2(N+1)} \sum_{k_1, k_2, k_3} I_{k, k_1, k_2, k_3} Q_{k_1} Q_{k_2} Q_{k_3}, \tag{1.28}$$

where the overlap integrals between the Fourier modes represent the momentum conservation:

$$I_{k, k_1, k_2, k_3} = \delta_{k - k_1 + k_2 - k_3, 0} + \delta_{k - k_1 - k_2 + k_3, 0}$$
$$-\delta_{k + k_1 + k_2 - k_3, 0} - \delta_{k + k_1 - k_2 + k_3, 0}. \tag{1.29}$$

In this case the energies of each mode $E_k$ turn to integrals of motion of the integrable limit. Upon deviation from the integrability the modes are intercoupled with a number of couplings scaling with the system size.

The classification proposed in Ref. [50] relies on the notion of the **coupling range** $\mathcal{R}$ of the network, i.e. *the number of inter-observable couplings*. The two limits presented above can now be classified in terms of their coupling range:

**Short-range network (SRN):** *The coupling range $\mathcal{R}$ is finite and independent from the number $N$ of degrees of freedom of the system.*

**Long-range network (LRN):** *The coupling range $\mathcal{R}$ increases with the number $N$ of degrees of freedom of the system, $\mathcal{R} = f(N)$, for a certain monotonically increasing function $f$.*

The former of the two presented regimes of KG Hamiltonian belongs to an SRN class. The Josephson junction chain model discussed above is another typical example. In this type of scenario the integrable limit is characterized by a set of **local observables (LOs)**, such as local energy, norm, charge, etc.

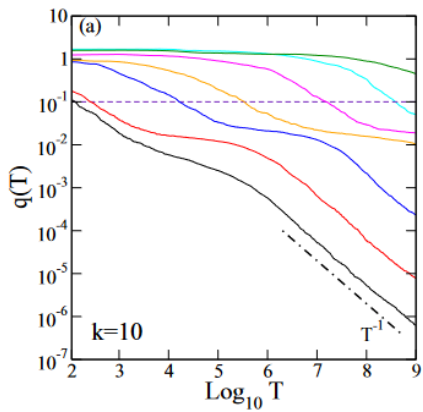

Figure 1.6: The dependence of the fluctuation index $q(T)$ corresponding to the set of finite time averages $\{\overline{E}_k\}$ in KG model as the system is tuned closer to the LRN integrable limit. The energy density varies from $h = 3.5$ (black) to $h = 0.0025$ (green). The wave number is fixed $k = 10$. The dashed line shows cutoff for measuring the ergodization time $T_E$. The figure is taken from Ref. [50].

The latter limit discussed in this section belongs to the LRN class. In general such cases are caused by the nonlinear perturbations resulting from approximations of two-body interactions. The perturbations are local in real space, but couple a large number of modes with each other due to the fact that the modes are extended, and thus are referred to as **extended observables (EOs)**.

Some models such as Klein-Gordon, Josephson junction chain, etc. may be tuned to both limits. Thus they are convenient testbeds for illustrating the framework. However this is not generally the case – the weakly nonlinear FPUT model belongs to the class of LRNs, however at variance to the KG there is no SRN limit of the FPUT model.

Short-range network class is characterized by dynamical glass regime with extremely large ergodization time to Lyapunov time ratio $T_E/T_\Lambda \gg 1$. What about LRNs? In Ref. [50] the analysis of both cases was performed using the KG Hamiltonian. The weak coupling $\varepsilon \to 0$ regime corresponding to SRN showed same results as the Josephson junction chain. The findings for the weak nonlinearity $h \to 0$ case corresponding to LRN were qualitatively different. In this case the set of energies of the normal modes $\{E_k\}$ for a fixed wavenumber $k$ was chosen for the analysis of the finite time averages (see Sec. 2.2 for details). In Fig. 1.6 from Ref. [50] the fluctuation index $q(T)$ is shown. Again as the system approaches the integrable limit (in this case LRN) the width of the distribution shows constant behavior before eventually decaying as expected due to ergodicity. In this case the equilibration time $T_E$ is measured at the cutoff value $q = 0.1$. Next the Lyapunov time $T_\Lambda$ is measured. For LRNs the result $T_E \sim T_\Lambda^2$ was found which is substantially distinct from SRNs. This result signifies of the existence of one relevant timescale – $T_\Lambda$.

### 1.4.3 Correlations and finite size effects

The fluctuation index $q(T)$ essentially corresponds to the width of the distribution of finite time averages $\varrho(T)$. The ergodic hypothesis suggests the decay of $q(T)$, which is in agreement with observations in both SRN and LRN (see Fig. 1.4 and Fig. 1.6). So far we have not discussed the decay law $q(T) \sim T^{-1}$. What are its origins? To shed some light on this matter the auto-correlation of the set of observables has to be taken into account. We will simplify the discussion by considering the second moment of the distribution $\mu_2(T)$. Over large time the first moment $\mu_1(T)$ approaches the phase space average, thus the decay of the fluctuation index $q(T)$ is essentially represented by the decay of $\mu_2(T)$. One may show that the second moment of the distribution is related to auto-correlation function $C(t) = \lim_{\tau \to \infty} \frac{1}{\tau} \int_0^\tau p_n(\tau) p_n(t + \tau) d\tau$ as $\mu_2(T) = \frac{1}{T} \int_0^T C(t) dt$, here $\{p_n\}$ is a set of observables under consideration [51]. Typically the correlations are assumed to decay exponentially fast in time. If so then the second moment $\mu_2$ and subsequently the fluctuation index $q(T)$ will decay as $q(T) \sim T^{-1}$.

In Ref. [51] these assumptions were tested for the Josephson junction chain model given by Eq. (1.23) in SRN limit $h \to \infty$. The second moment $\mu_2(T)$ was computed for the set of averages of momenta $\{p_n\}$ and varying system size. The result (see Fig. 1.7) revealed a new system size dependent timescale $T_D$ up to which the second moment shows decay $\mu_2(T) \sim T^{-1/2}$. For times $T > T_D$ there exists a transition to $\mu_2(T) \sim T^{-1}$. It follows that up to times $T_D$ the correlation function $C(t)$ does not decay exponentially but according to the power law $C(t) \sim t^{-1/2}$. The new timescale increases with the system size. Thus it follows that in the infinite size limit limit $T_D(N \to \infty) \to \infty$ and thus $\mu_2(T) \sim T^{-1/2}$ in the infinite systems is expected. After rescaling the curves presented in Fig. 1.7 the dependence of the new timescale on the system size was obtained $T_D \sim N^2$ signifying of a certain diffusion law. Before proceeding we note that the results on the fluctuation index $q(T)$ in Fig. 1.4 in principle should also display similar behavior. Inability to observe it is attributed to large fluctuations.

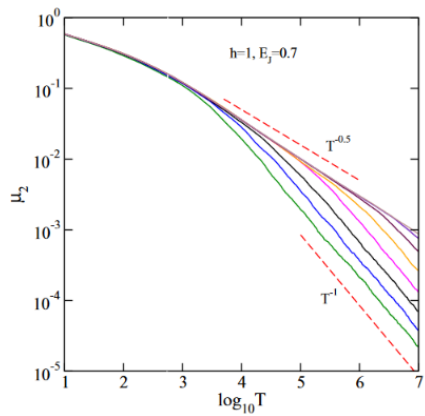

Figure 1.7: The dependence of the second moment $\mu_2(T)$ corresponding to the set of finite time averages $\{\bar{p}_n\}$ in Josephson junction chain model as the system size is varied from $N = 2^6$ (bottom) to $N = 2^{13}$ (top). The energy density and coupling are fixed $h = 1$, $E_J = 0.7$. The figure is taken from Ref. [51].

The origins of the new timescale take root from presence and propagation of chaotic resonances in the chain similarly to what was discussed in Sec. 1.4.1. The condition for resonance is given by $\Delta+_n < E_J$ and $\Delta_n^- < E_J$ where $\Delta_n^\pm = |p_n(p_n - p_{n\pm1})|$. In Ref. [51] the probability of resonance was computed and is

given by:
$$\pi_r \approx \frac{1}{4}\left(\frac{E_J}{h}\right)^2. \qquad (1.30)$$

It follows that the distance between chaotic triplets $p_{n-1}, p_n, p_{n+1}$ grows as $l_r = (h/E_J)^2$ in proximity to the SRN integrable limit with decreasing $E_J$. Thus effectively the whole system is subdivided into $N/l_r$ patches of length $l_r$ with one resonant triplet per patch. The equilibration starts with spreading of the resonance through the patch. Before the time $T_E$ the thermalization of actions involved in the resonances is not noticeable on the scale of the whole system. Thus the width of the corresponding finite time averages distributions stays constant. When the spread of resonances becomes comparable to that of the length of the patch one observes the onset of ergodization. The dynamics in the system is diffusive thus the estimated timescale is $T_E \sim \sqrt{Dl_r}$ where $D$ is the diffusion coefficient. The resonances diffuse until the whole patch of $l_r$ is equilibrated, which happens for $T_D \sim N^2$ [51, 62]. Until this time the width of finite time averages distributions is observed to decay as $\mu_2(T) \sim T^{-1/2}$. For times $T > T_D$ the typical $\mu_2(T) \sim T^{-1}$ is observed, i.e. the whole system has equilibrated.

### 1.4.4 Open questions

The main progress attributed to the recent studies is the identification of the range of networks of observables as a property significantly affecting the ergodization dynamics. The properties characteristic to SRNs and LRNs have been identified and studied to a relative extent. At the same time one could pose a reverse question: given the dynamics of some observables can one identify the class of the network coupling them? Frequently the models describing the observed dynamics is not known, but the raw data such as the time evolution of coordinates is available. Can we make predictions based on that data? Answering such questions runs into several problems.

First we note that in the sections above the Hamiltonian structure allowed for identification of integrable limits and the corresponding actions. However it is not guaranteed that such identification is possible. In some cases the choice of actions may be ambiguous. Thus the FPUT model may be viewed as the harmonic chain perturbed by nonlinear couplings; but alternatively it may be viewed as a perturbation of another integrable model – the Toda chain [66, 67]:

$$H_{\text{Toda}} = \sum_n \left[\frac{1}{2}p_n^2 + e^{-(q_{n+1}-q_n)} + (q_{n+1} - q_n) - 1\right]. \quad (1.31)$$

In the limit of small energy density, the larger powers of the exponential term are negligible. In this manner the FPUT model may be viewed as the nonintegrable perturbation to the Toda model [68].

The picture turns grimmer given that even the conclusions regarding the chaoticity and nonintegrability cannot be reached in a general case scenario. In particular for certain choice of observables integrable setups such as harmonic oscillator chains and the Toda chain may show ergodic, thermal-like behavior [55, 28]. In other cases the observed ergodization times in weakly nonintegrable systems such as FPUT may vary if instead of normal modes the local observables are chosen [48].

Clearly the observable choice is an issue which may lead to ambiguous results. Thus to answer the questions posed above we have two ways: 1) Perform a careful investigation of observable dynamics with the goal on identifying tools for inferring the class of nonintegrable dynamics; 2) Find observable choice independent tools for classification of weakly nonintegrable dynamics.

Some other intriguing questions include the search for other classes of nonintegrable networks. To this end the models with exponentially or polynomially decaying couplings are of an interest as well as disordered models with nonlinearity.

## 1.5 Outline

In the next chapter we address the questions discussed above using several approaches and techniques.

In Sec 2.2 we focus on the dynamics of observables and characterize thermalization using the finite time averages. We follow the dynamics of local and extended observables in weakly nonintegrable SRNs and LRNs with the goal to identify the differences in their dynamics and explore the consequences of the choice of observables. We extract thermalization timescale in line with the approach introduced in Sec. 1.3.1 and compare to those given by Lyapunov characteristic exponent computation. We further study the scaling of the size of the tangent vectors with the system size to characterize how chaos spreads in the system depending on the network range.

In Sec. 2.3 we offer a framework of identification of the classes of weakly nonintegrable dynamics while avoiding the ambiguity related to observable choice in terms of the whole Lyapunov spectrum. We compute the spectra in LRNs and SRNs and find significant qualitative difference. In Sec. 2.4 we follow up with the application of this approach to the disordered systems as well as the Hamiltonian setups such as the Klein-Gordon model discussed above.

Investigating the dynamics in extreme proximity to integrable limit is a challenging task. Computing the finite time averages requires simulations of a large number of trajectories simultaneously. The computation of full Lyapunov spectra is a numerically challenging task as the evolution of tangent vectors is done together with the reference trajectory. Time continuous Hamiltonian dynamics is inefficient due to errors introduced by time discretization which affects the fulfillment of conservation laws. Numerical integration of time-continuous dynamics requires implementation of symplectic integrator schemes which limits numerics from the perspective of both evolution time and system size. To overcome these difficulties we use a novel tool – unitary maps. Such models possess all necessary features of weakly nonintegrable dynamics of Hamiltonian systems such as phase space area preserving evolution, conservation of macroscopic quantity such as energy, etc. We start the next chapter by introducing the model and elaborating on its effectiveness for simulating physical phenomena.

# Chapter 2

# Study of thermalization dynamics of macroscopic weakly nonintegrable systems using unitary maps

## 2.1 Unitary maps

Our goal is to describe the time scales associated with equilibration of weakly nonintegrable systems. There exist several numerical challenges associated with this objective: 1) The thermalization time scales are expected to grow quickly with increasing proximity to the integrable limit. 2) Describing large scale systems is numerically demanding. 3) The continuous time evolution of typical Hamiltonian setups requires error control which puts additional constraints on the effectiveness of numerical simulations. Thus, there exists a demand for numerically efficient models which show representative dynamics. One possible solution to answer this call is to use discrete-time maps instead of continuous-time Hamiltonian dynamics. Thus the errors associated with the discretization of time are avoided and all conservation laws are satisfied with the best possible precision. Using maps to illustrate the principal properties of chaotic dynamics is a longstanding tradition in the field of dynamical systems. Some famous examples include Arnold cat map [19], Chirikov standard map [69], logistic map [70], etc. The unitary map evolution protocols discussed below allow for efficient computations while possessing the necessary properties for simulating a rich variety of physical phenomena including Anderson localization [71, 72, 73], breather solutions [74], molecular dynamics [75], topological states of matter [76], etc.

One of the renowned proposals to use unitary evolution maps to simulate physical processes was by Aharonov, et al. with the idea to quantize the classical random walk process [77]. Instead of tossing a coin to determine a subsequent position of a walker one now introduces a "quantum coin" in terms of a unitary operator – a map from the initial state at time $t$ to as state $t+1$. The resulting evolution of the walker (now a wave function) significantly differs from its classical counterpart. A well-known example is related to the return probability $p$. For for classical random walk in $1D$ and $2D$ the probability of the walker to return to the origin is $p = 1$, while for quantum random walk the return probability is $p < 1$ even in $1D$ [78]. Another fundamental difference is related the spreading of a localized state, which is diffusive in case of classical walks, but ballistic for quantum case. This property has been used extensively for quantum computing algorithms development, where quantum walks were implemented in search algorithms [79, 80, 81].

The unitary circuit maps discussed in the current thesis are a generalization of discrete time quantum walks inspired by quantum unitary circuits [82]. The basic idea lies in coupling local Hilbert spaces with dimensionality $q$ via unitary transformations performed using gate operations. In modern physics unitary circuits are used as a setup for studies of information scrambling and strongly coupled systems [83, 84, 85]. At the same time, unitary circuits serve as a model to simulate quantum chaos [86, 87].

In what follows we consider a simplified model with $q = 1$ local degree of freedom, i.e. one complex component per site. We introduce nonlinearity in analogy with mean-field Hamiltonian terms which makes the dynamics classical. We then study the ergodicity and chaotic properties of the system by integrating the trajectories in the corresponding phase space.

There are several motivational points for using unitary circuit maps to simulate physical phenomena. The main reason lies in considerable computational advantages compared to Hamiltonian setups. The discrete-time evolution nature of the map allows for speedups up to several orders in magnitude [88]. We elaborate on the numerical efficiency of unitary maps in Sec. 2.1.5. On the other hand unitary maps allow for novel physical regimes, such as Anderson localization with unique localization length for all eigenstates [71]. In the current work we expand the range of possible applications of unitary circuit maps by performing a comprehensive study of weakly nonintegrable dynamics in terms of statistics of the finite time averages of observables (see Sec. 2.2) and the Lyapunov characteristic exponents (see Sec. 2.3).

### 2.1.1 Model

The unitary circuit map model consists of one-dimensional chain of $N/2$ unit cells characterized by a complex valued vector of size $N$ containing all the information regarding unit cell sites $A$ and $B$ (see Fig. 2.1):

$$\vec{\Psi}(t) = \left(\psi_1^A(t), \psi_1^B(t), \psi_2^A(t), \psi_2^B(t) \dots \psi_{N/2}^A(t), \psi_{N/2}^B(t)\right). \quad (2.1)$$

The nonlinear evolution of the system is governed by a discrete unitary map consisting of several transformations of the vector $\vec{\Psi}$:

$$\hat{U} = \sum_n \hat{G}_n \sum_n \hat{C}_{B,A} \sum_n \hat{C}_{A,B}. \quad (2.2)$$

where $\hat{C}_{A,B}$ and $\hat{C}_{B,A}$ are given by unitary matrices coupling the neighboring sites:

$$\sum_n \hat{C}_{A,B}\vec{\Psi}(t) = \sum_n \begin{pmatrix} \cos\theta & \sin\theta \\ -\sin\theta & \cos\theta \end{pmatrix} \begin{pmatrix} \psi_n^A(t) \\ \psi_n^B(t) \end{pmatrix},$$

$$\sum_n \hat{C}_{B,A}\vec{\Psi}(t) = \sum_n \begin{pmatrix} \cos\theta & \sin\theta \\ -\sin\theta & \cos\theta \end{pmatrix} \begin{pmatrix} \psi_n^B(t) \\ \psi_{n+1}^A(t) \end{pmatrix}.$$

$$(2.3)$$

The local transformations $\hat{C}$ can be in general represented by an arbitrary $2 \times 2$ unitary matrices. Our particular choice is parametrized by a single angle $\theta$ which plays the role of a hopping parameter strength in Hamiltonian systems by coupling the nearest neighbor sites. In principle the evolution protocol using unitary matrices of some generic dimension $d \times d$ to couple $d$ neighboring sites can be constructed and will immediately result in a multiple-band problem. We follow the simplest and most frequently used unitary map protocol with $2 \times 2$ matrices and thus two components per unit cell and a two-band model [82, 77, 89, 88]. We would like to point out that the physics in the following sections is independent from the number of bands and has been verified (although on smaller timescales and system sizes) for the case of single-band Hamiltonian dynamics.

The transformation $\hat{G}_n$ induces a nonlinear factor $g|\psi_n(t)|^2$ is an analog of a mean-field potential:

$$\hat{G}_n \psi_n^{A,B}(t) = e^{ig|\psi_n^{A,B}(t)|^2} \psi_n^{A,B}(t). \tag{2.4}$$

The trajectories of the unitary circuit map evolve in a $2N$ dimensional phase space as each complex component $\psi_n^{A,B}$ is parametrized by two real-valued numbers. In this sense one is able to follow a trajectory in a similar fashion to the classical Hamiltonian [50, 49, 51] or map dynamics. Due to the unitary evolution of the map, the total norm of the state vector is conserved over time $|\vec{\Psi}(t)| = const.$ This is analogous to energy conservation in Hamiltonian dynamics.

### 2.1.2 Equations of motion and diagonalization

The evolution of the state vector components $\vec{\Psi}(t)$ obeys the following equations of motion:

$$\psi_n^A(t+1) = e^{ig|\varphi_n^A(t)|^2} \varphi_n^A(t) \quad , \quad \psi_n^B(t+1) = e^{ig|\varphi_n^B(t)|^2} \varphi_n^B(t). \tag{2.5}$$

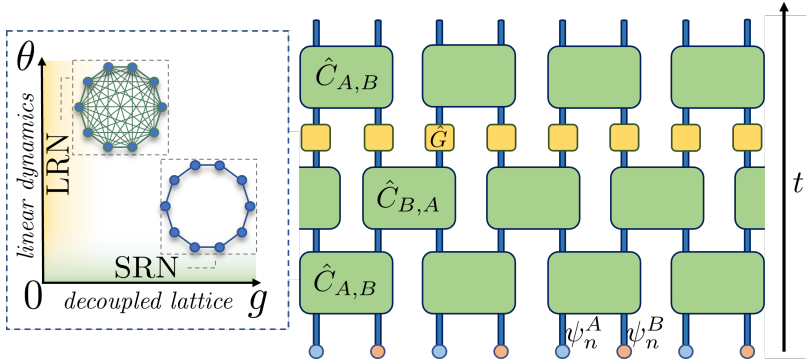

Figure 2.1: A schematic representation of the unitary circuits map and the parameter space $\{g, \theta\}$. The black arrow on the right indicates the flow of time. The state is represented by blue and pink dots for $\psi_n^A$ and $\psi_n^B$ respectively, with subsequent applications of unitary transformations $\hat{C}_{A,B}$ and $\hat{C}_{B,A}$ (large green blocks) parametrized by the angle $\theta$, and local nonlinearity generating maps $\hat{G}$ (small yellow blocks) parametrized by the nonlinearity strength coefficient $g$. In the parameter space the highlighted areas correspond to the networks of coupled actions induced by respective weak nonintegrable perturbation. Integrable limits are reached for $g = 0$ (linear evolution of extended normal modes) or $\theta = 0$ (decoupled nonlinear map lattice). Small nonzero $g$ values induce LRNs, small nonzero $\theta$ values induce SRNs. The network images indicate actions (filled circles) coupled due to nonintegrable perturbation (straight lines).

where $\varphi_n^{A,B}(t)$ are the components of the state vector $\vec{\Psi}$ after the application of the mixing maps $\hat{C}_{A,B}$ and $C_{B,A}$ (see Fig. 2.1):

$$\varphi_n^A(t) = \left[\cos^2\theta \psi_n^A(t) - \cos\theta\sin\theta \psi_{n-1}^B(t) + \sin^2\theta \psi_{n+1}^A(t) + \cos\theta\sin\theta \psi_n^B(t)\right],$$

$$\varphi_n^B(t) = \left[\sin^2\theta \psi_{n-1}^B(t) - \cos\theta\sin\theta \psi_n^A(t) + \cos^2\theta \psi_n^B(t) + \cos\theta\sin\theta \psi_{n+1}^A(t)\right]. \tag{2.6}$$

In the linear case $g = 0$ the solution can be determined exactly using the standard ansatz $\left(\psi_n^A(t), \psi_n^B(t)\right)^T = e^{-i(\omega_k t - kn)} \left(\psi_k^A, \psi_k^B\right)^T$ with eigenfrequencies $\omega_k$ and wave numbers $k$. The dispersion relation for the eigenfrequencies $\omega(k)$ is determined from Eq. (2.5):

$$\omega(k) = \pm \arccos\left(\cos^2\theta + \sin^2\theta \cos k\right), \tag{2.7}$$

with two dispersive bands $\omega_k^\alpha$ ($\alpha = 1, 2$). The corresponding eigenvectors in $k$ space have two components which obey the following relation:

$$\frac{\psi_k^{\alpha,A}}{\psi_k^{\alpha,B}} = \frac{\cos^2\theta + \sin^2\theta e^{ik} - e^{i\omega_k^\alpha}}{\sin\theta \cos\theta(1 - e^{ik})}. \tag{2.8}$$

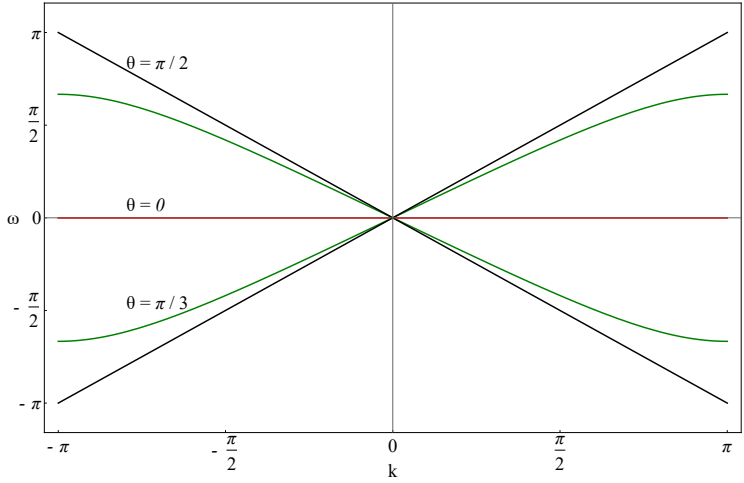

Figure 2.2: The dispersion relation $\omega(k)$ for different values of $\theta$: $\theta = 0$ (flat band, red solid line), $\pi/3$ (green line), $\pi/2$ (linear dispersion black line).

The normal modes are then given as:

$$\vec{\Psi}_k^\alpha = \sum_n e^{ikn} \vec{\psi}_k^\alpha, \quad \vec{\psi}_k^\alpha = \left(\psi_k^{\alpha,A}, \psi_k^{\alpha,B}\right)^T. \tag{2.9}$$

The parameter $\theta$ controls the width of each band. For a generic value of $\theta$ the two bands have finite width (e.g. green lines, $\theta = \pi/3$ in Fig. 2.2). For $\theta = \pi/2$ the two bands turn into straight lines, a one-dimensional Dirac-like cone (black lines in Fig. 2.2). Finally for $\theta = 0$ the spectrum $\omega(k) = 0$ turns to a flat band regime (red lines in Fig. 2.2), which corresponds to a macroscopic degeneracy. Note that in our model matrices involved in local transformations $\hat{C}_{A,B}$ and $\hat{C}_{B,A}$ are the same, which results in a gapless spectrum. To open the gap in the spectrum one would need to introduce two parameters $\theta_1$ and $\theta_2$, in maps $\hat{C}_{A,B}$ and $\hat{C}_{B,A}$. However, such cases are beyond the scope of our discussion.

### 2.1.3 Integrable limits

We consider two integrable limits in the parameter space $\{g,\ \theta\}$ of the unitary circuits map: vanishing coupling: $\{\theta = 0,\ g \neq 0\}$, and vanishing nonlinearity: $\{g = 0,\ \theta \neq 0\}$ (see Fig. 2.1). In each case the integrable limit is characterized by a corresponding set of $N$ observables (actions) which are decoupled and thus are conserved in time during the evolution. In what follows we investigate each of the integrable limits in detail by deviating from it slightly and thus inducing a coupling between the actions. The obtained networks of observables are classified in accord with the discussion in Sec. 1.4.2.

#### 2.1.3.1 Short-range network

In the limit of vanishing coupling $\theta = 0$ the local transformations $\hat{C}$ turn into identity matrices resulting in a trivial evolution of the state components with a trivial accumulation of phase (see Eq. (2.5) for details):

$$\psi_n^{A,B}(t) = e^{ig|\psi_n^{A,B}(t)|^2 t}\psi_n^{A,B}(0). \qquad (2.10)$$

The amplitudes $|\psi_n^{A,B}|$ are time independent and conserved in the integrable limit. Introducing a small value of $\theta \neq 0$ results in coupling of the amplitudes. Approx-imating equations of motion (2.5) for small values of $\theta$ we obtain:

$$\psi_n^A(t+1) = e^{ig|\varphi_n^A(t)|^2}\varphi_n^A(t),$$
$$\varphi_n^A(t) = \left[\psi_n^A(t) - \theta(\psi_{n-1}^B(t) - \psi_n^B(t))\right],$$

$$\psi_n^B(t+1) = e^{ig|\varphi_n^B(t)|^2}\varphi_n^B(t),$$
$$\varphi_n^B(t) = \left[\psi_n^B(t) + \theta(\psi_{n+1}^A(t) - \psi_n^A(t))\right].$$
(2.11)

These equations of motion couple the observables through nearest neighbor terms and fall under the definition of a short-range network (see Fig. 2.1).

#### 2.1.3.2 Long-range network

Generally, a state vector $\vec{\Psi}(t)$ may be decomposed in terms of normal modes of the linear system (see Eq. (2.9)):

$$\vec{\Psi}(t) = \sum_{k,\alpha} c_k^\alpha(t)\vec{\Psi}_k^\alpha. \tag{2.12}$$

The coefficients $c_k^\alpha$ are complex-valued numbers. In the linear regime $g = 0$ the evolution of the normal mode coefficients is given by the phase rotation $c_k^\alpha(t) = c_k^\alpha(0)e^{i\omega_k^\alpha t}$, and the norm $|c_k^\alpha|$ is conserved over time. Introducing a small but nonzero value of $g \neq 0$ results in coupling of the modes. Approximating equations of motion (2.5) for small values of $g$ we obtain:

$$c_k^\alpha(t+1) = e^{i\omega_k}c_k^\alpha(t) +$$

$$\tfrac{ig}{N}\sum_{\substack{\alpha_1,\alpha_2,\alpha_3\\k_1,k_2,k_3}} e^{i(\omega_{k_1}^{\alpha_1}+\omega_{k_2}^{\alpha_2}-\omega_{k_3}^{\alpha_3})} I_{k,k_1,k_2,k_3}^{\alpha,\alpha_1,\alpha_2,\alpha_3} c_{k_1}^{\alpha_1}(t)c_{k_2}^{\alpha_2}(t)\left(c_{k_3}^{\alpha_3}(t)\right)^*, \tag{2.13}$$

$$I_{k,k_1,k_2,k_3}^{\alpha,\alpha_1,\alpha_2,\alpha_3} = \delta_{k_1+k_2-k_3-k,0}\sum_p \psi_{k_1}^{\alpha_1,p}\psi_{k_2}^{\alpha_2,p}(\psi_{k_3}^{\alpha_3,p})^*(\psi_k^{\alpha,p})^*. \tag{2.14}$$

All $c_k^\alpha$ are intercoupled according to the second term in Eq. (2.13). For each action $c_k^\alpha$ the number of elements in the sum is proportional to $N^2$ due to the constraints enforced by the overlap integrals in Eq. (2.14). This case falls under the definition of a long-range network.

### 2.1.4 Tangent vectors and their dynamics

In what follows we compute the Lyapunov time which requires the simultaneous evolution of a reference trajectory together with the tangent vector (see Sec. 1.3.2). To derive the equations of motion for tangent vectors we decompose a trajectory $\vec{\Psi}(t)$ as an unperturbed trajectory $\vec{x}$ and a deviation $\vec{w}$:

$$\vec{\Psi}_i(t) = \vec{x}(t) + \vec{w}(t). \tag{2.15}$$

For convenience we redefine the linear part of evolution in terms of functions $\alpha_n^{A,B}[\vec{\Psi}(t)]$:

$$\alpha_n^A[\vec{\Psi}(t)] \equiv \cos^2\theta \psi_n^A(t) - \cos\theta\sin\theta \psi_{n-1}^B(t) + \sin^2\theta \psi_{n+1}^A(t) + \cos\theta\sin\theta \psi_n^B(t)$$

$$\alpha_n^B[\vec{\Psi}(t)] \equiv \sin^2\theta \psi_{n-1}^B(t) - \cos\theta\sin\theta \psi_n^A(t) + \cos^2\theta \psi_n^B(t) + \cos\theta\sin\theta \psi_{n+1}^A(t).$$
$$\tag{2.16}$$

Now together with the nonlinear part the equations of motion can be written as:

$$\psi_n^A(t+1) = e^{ig|\alpha_n^A[\vec{x}(t)+\vec{w}(t)]|^2} \alpha_n^A[(\vec{x}(t)+\vec{w}(t))]$$
$$\psi_n^B(t+1) = e^{ig|\alpha_n^B[\vec{x}(t)+\vec{w}(t)]|^2} \alpha_n^B[(\vec{x}(t)+\vec{w}(t))].$$
$$\tag{2.17}$$

Expanding the nonlinear term and keeping terms only to the first order of $\vec{w}$ results in

$$|\alpha_n^p[\vec{x}(t) + \vec{w}(t)]|^2 = |\alpha_n^p[\vec{x}(t)] + \alpha_n^p[\vec{w}(t)]|^2 =$$
$$\alpha_n^p[\vec{x}(t)]\alpha_n^p[\vec{x}(t)]^* + \alpha_n^p[\vec{w}(t)][\alpha_n^p[\vec{w}(t)]^* +$$
$$\alpha_n^p[\vec{w}(t)]\alpha_n^p[\vec{x}(t)]^* + \alpha_n^p[\vec{x}(t)]\alpha_n^p[\vec{w}(t)]^* \approx$$
$$|\alpha_n^p[\vec{x}(t)]|^2 + \Delta_n^p(t), \tag{2.18}$$

where

$$\Delta_n^p(t) = \alpha_n^p[\vec{x}(t)]\alpha_n^p[\vec{w}(t)]^* + c.c. \tag{2.19}$$

Thus we can rewrite the exponential term in Eq. (2.17):

$$e^{ig|\alpha_n^p[\vec{x}(t)+\vec{\varepsilon}(t)]|^2} = e^{ig|\alpha_n^p[\vec{x}(t)]|^2}\left[1 + ig\Delta_n^p(t)\right], \tag{2.20}$$

and using the linearity of $\alpha_n^p[\vec{\Psi}(t)]$ we finally arrive at the following linear equations:

$$w_n^p(t+1) = e^{ig|\alpha_n^p[\vec{x}(t)]|^2}\left\{\alpha_n^p[\vec{w}(t)] + ig\Delta_n^p(t)\alpha_n^p[\vec{x}(t)]\right\}. \tag{2.21}$$

## 2.1.5 Computational effectiveness of Unitary Maps

In the following sections we aim to use the unitary maps as a computational tool for studying the ergodization and mixing properties of weakly non-integrable dynamics. The discrete-time nature of the evolution is the key feature that allows us to extend the simulation times beyond the limits reachable for Hamiltonian dynamics. The local maps $\hat{C}$ act independently on different parts of the system. This allows us to use the parallelization technique with modern multi-core processors. The evolution protocol consists of cheap arithmetic operations and does not require the precalculation of integrating coefficients which is frequent in Runge-Kutta as well as symplectic integration methods.

The other significant advantage comes with the lack of errors associated with the conservation of macroscopic state functions such as total energy, momentum, etc. The disadvantage of time-continuous Hamiltonian evolution frequently comes with a dilemma – one either chooses fast integration time while sacrificing the accuracy, or vice versa one is able to achieve accurate results, but within limited time scales. This dilemma is completely resolved when one considers unitary map evolution. The conserved quantities, such as the norm are conserved exactly, while the integration time is extended to several more orders of magnitude than allowed by Hamiltonian setups. This point is well demonstrated by recent studies of nonlinear wave-packet spreading with evolution times of the order $t = 2 \cdot 10^{12}$, with accumulated error of the order of $10^{-4}$, which comes from the machine's round-off error[88].

## 2.2 Observable dynamics in macroscopic weakly non-integrable systems

In the Sec. 1.4.4 we discussed the issue of the choice of observables. Typically the studies on thermalization of dynamical systems focus on a specific choice and once fixed, the universality of the results from other standpoints is rarely questioned. In this manner the integrable setups may show seeming thermalization [28, 55] due to ergodicity on the tori. Despite the infinite number of possible observable choices, two are dominating: *local observables (LOs)* (local norm, charge, energy, etc) and *extended observables (EOs)* (normal modes). In the Sec. 1.4.2 we elaborated on these choices and the integrable setups where they turn into the actions of the short-range and long-range integrable limits respectively.

Our goal is to cover a broad range of weakly nonintegrable setups and observable choices by classifying the ergodic properties of LOs and EOs in both SRN and LRN settings. The unitary circuit map allows us to consider extreme proximity to the integrable limit in systems with the size up to $N = 10^5$ and time scales up to $T_{max} = 10^9$. We demonstrate that this approach succeeds in a full and unique characterization of the thermalization dynamics despite possible observable choice ambiguity [90]. We compute the thermalization times in line with the approach discussed in the Sec. 1.3.1. As a comparative timescale we compute the Lyapunov time $T_\Lambda = 1/\Lambda_{max}$ and find qualitatively distinct behavior for SRNs and LRNs. The following sections are based on the Ref. [90].

### 2.2.1 Methods

#### 2.2.1.1 Finite time averages

Our goal is to study the ergodization of local and extended observables in two distinct integrable limits. Ergodization is a process of time averages of observables approaching their phase space averages. In line with this definition we define sets of finite time averages of observables in SRN and LRN cases and study the statistical properties of these sets, in particular their variance. For a time-dependent observable $o(t)$ we define a finite time average as:

$$\bar{o}_T = \frac{1}{T} \sum_{t=0}^{t=T} o(t) \qquad (2.22)$$

We construct a set of finite time averages $\{\bar{o}_T\}_M$ by following $M$ trajectories. This set is characterized by a distribution with probability density function $\varrho(\bar{o}_T)$. From the ergodization hypothesis we expect each $\bar{o}_{T\to\infty} = \langle o \rangle$, where $\langle o \rangle$ is an average taken over the phase space. The distribution $\varrho$ is therefore expected to peak around the phase space average, reducing its variance to zero and thus approaching a delta-function for infinite averaging times: $\varrho(\bar{o}_{T\to\infty}) = \delta(\bar{o}_{T\to\infty} - \langle o \rangle)$. We study the convergence by following the variance $\sigma^2(T)$ of the distribution $\varrho(\bar{o}_T)$. We perform the analysis for two distinct observables corresponding to conserved quantities in the integrable limit – local observables $|\psi_n|$, and extended observables $|c_k^\alpha|$ which are the amplitudes of normal modes coefficients.

### 2.2.1.2 Lyapunov time and tangent vector

Typically the equilibration time scales are compared to intrinsic time scales of chaotic dynamics [49, 50]. The shortest timescale of chaotic motion is given in terms of the Lyapunov time $T_\Lambda$. We follow the approach discussed in Sec. 1.3.2 and characterize chaoticity related timescale of non-integrable dynamics. We compute the largest Lyapunov exponent and characterize the corresponding tangent vector. At any given time the normalized tangent vector points in the direction of the strongest chaotization. We compute the time average of the participation number $\overline{PN}$ of the normalized tangent vector $\mathbf{w}(t)$:

$$PN(t) = \frac{1}{\sum_i |w_i(t)|^4}. \qquad (2.23)$$

The tangent vector depends on the coordinate basis choice. Once the basis is fixed, a delocalized tangent vector results in $PN \sim N$ while a localized tangent vector yields a system size independent value of $PN$.

### 2.2.1.3 Evolution and initial conditions

We perform the evolution of the state vector $\vec{\Psi}(t)$ in coordinate space using the unitary map defined in Eq. (2.5). Before measuring the observables the system is prerun to ensure thermalization. The initial conditions for each component of the state vector are set as $\psi_n^{A,B} = r_n e^{i\gamma_n}$. For each of the $M$ initial conditions (trajectories) we generate the amplitudes $r_n$ as uncorrelated random numbers to be distributed according to the distribution with probability density function $f(x) \sim x e^{-x^2}$ in accord with the Gibbs distribution for the norm densities. We then renormalize the state vector such that the norm density is set to unity. The phases $\gamma_n$ are distributed uniformly on the interval $[0, 2\pi]$.

### 2.2.1.4 Observables

**LOs:** Due to the translational invariance in the system, we assume the local observables (LOs) to be statistically identical and independent. This allows to generate a set of finite time averages of LOs from a single trajectory. This way an output of a single run will generate the variance $\sigma^2(T)$ of a set of finite time averages of $N$ observables.

**EOs:** In contrast to LOs the normal mode coefficients $c_k^\alpha$ - which are the extended observables (EOs) - are not statistically identical for different values of $k$. Thus we choose a specific value of the wave vector $k = \pi/2$ and perform $M$ trajectory runs. We extract the time average $\overline{c}_k^m(T)$ to obtain the set of $M$ finite time averages $\{\overline{c}_k^m(T)\}_{m=1}^M$ whose variance $\sigma^2(T)$ is then computed.

### 2.2.1.5 Time scales

For the sake of clarity we briefly list and remind the time scales involved in our studies. The ergodization time $T_E$ is the timescale up to which the observables which turn integrals of motion at the very integrable limit stay essentially constant for a weakly nonintegrable system. It follows that $T_E$ is diverging upon approaching the integrable limit. We study nonlinear systems with short range coupling in real space, which allow for any of the nonintegrable network ranges (short and long), as the latter are defined in the corresponding space of actions which turn integrals of motion at the very integrable limit. Therefore, regardless of the type of nonintegrable network, once the timescale $T > T_E$ is reached, diffusion in real space sets in. A second diffusion time $T_D \sim N^2$ marks the time at which the diffusion in real space reaches the boundaries of the finite system. Note that for an infinite system size $T_D$ diverges for any weakly nonintegrable system which still has a finite (though potentially very large) ergodization time $T_E$. Finally, we compare the above time scales with the Lyapunov time $T_\Lambda = 1/\Lambda_{max}$ which is given by the inverse of the largest Lyapunov exponent $\Lambda_{max}$. In all cases $T_\Lambda \leq T_E$ which is an expected result - there can be no ergodization and thermalization before any chaos sets in. However, we find that the ratio $T_E/T_\Lambda$ diverges much faster with increasing $T_E$ for SRNs, while stays almost constant for LRNs.

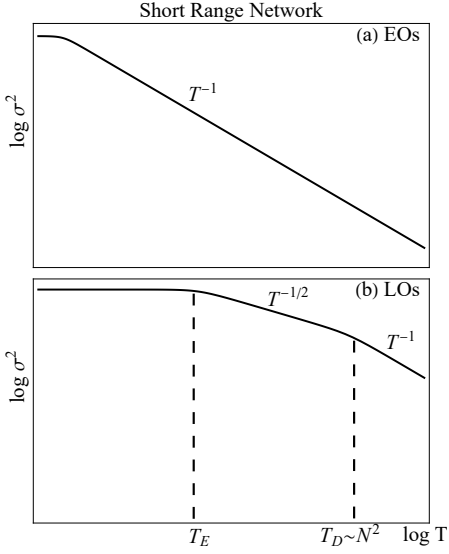 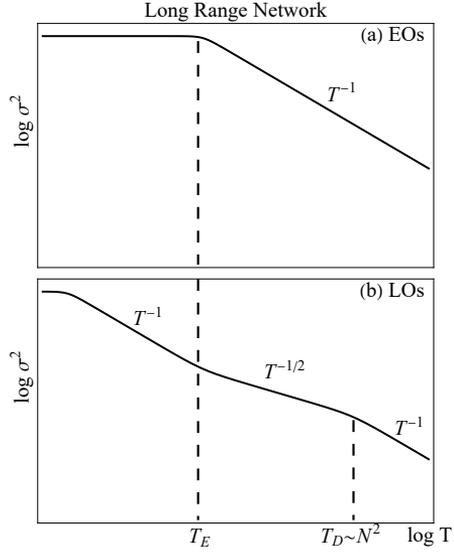

Figure 2.3: Expected results in SRN case (see Sec. 2.2.2).

Figure 2.4: Expected results in LRN case (see Sec. 2.2.2).

### 2.2.2 Expected Results

#### 2.2.2.1 Short-range network

In the integrable limit $\theta = 0$ the LOs are decoupled and the variance of the set of finite time averages of LOs will stay constant over time $\sigma^2(T) = const$. Once the small deviation $\theta \neq 0$ has been introduced the LOs are coupled and the dynamics shows nonintegrable behavior. This is the regime of dynamical glass discussed in Sec. 1.4.1 The weak coupling of LOs will manifest in nearly frozen actions with rare resonant spots in the system, where chaotic dynamics takes place [49]. The mean distance between the chaotic spots grows upon approaching the integrable limit. The strength of chaotic dynamics (largest Lyapunov exponent) diminishes upon approaching the integrable limit. This interplay between chaotic and non-chaotic parts of the system will result in the ergodization time $T_E$ - a timescale on which the chaotic spots diffuse over a distance of the order of the average spacing between the resonances. It follows that $\sigma^2(T < T_E)$ stays approximately constant up to $T_E$. That timescale $T_E$ will diverge upon approaching the integrable limit. At finite distance from the integrable limit the resonances continue to diffuse through the system resulting in $\sigma^2(T) \sim T^{-1/2}$ for $T_E < T < T_D$ [51].

Once the excitations diffuse across the entire system all correlations vanish and we expect $\sigma^2(T) \sim T^{-1}$ due to finite size effects. The timescale of the transition from $\sigma^2(T) \sim T^{-1/2}$ to $\sigma^2(T) \sim T^{-1}$ is denoted as $T_D \sim N^2$ [51]. This behavior is in line with the findings discussed in Sec. 1.4.3

The EOs are not conserved even at the very integrable limit. They show fast fluctuations and quick pseudo-thermalization in analogy with Ref. [55]. Thus we expect an immediate $\sigma^2(T) \sim T^{-1}$ decay starting from the shortest time scales.

We provide a schematic representation of the behavior of the variance $\sigma^2(T)$ versus time in log-log scale in Fig. 2.3.

#### 2.2.2.2 Long-range network

In the integrable limit $g = 0$ the system dynamics is linear, and we expect $\sigma^2(T) = const$ for EOs. Upon the deviation from the limit the variance is expected to decay $\sigma(T)^2 \sim T^{-1}$ after some ergodization time $T_E$ required for spread of chaos into the network (see Fig. 2.4). We expect time $T_E$ to be of the order of Lyapunov time $T_\Lambda$ as there appears to be no other timescale governing the dynamics.

LOs will show a more involved outcome. First, even at the integrable limit they are not conserved, and show fast fluctuations and quick seeming thermalization in analogy with Ref.[55]. Thus we expect an immediate $\sigma^2(T) \sim T^{-1}$ decay starting from the shortest time scales. However, this only holds up to $T_E$ if the system is large enough. Indeed, the EOs are preserved up to $T_E$ and correspond to ballistically propagating modes (waves) in real space having a largest finite group velocity $v_g$. If the system size $N \gg v_g T_E$ the modes will start to interact and ballistic propagation is replaced by diffusive propagation in real space for $T > T_E$. Correspondingly the LO dynamics results in a crossover from $\sigma^2(T) \sim T^{-1}$ to $\sigma^2(T) \sim T^{-1/2}$ for $T > T_E$. At a timescale $T_D \sim N^2$ the diffusion reaches the system boundaries, and the LO dynamics crosses over back to a final asymptotic $\sigma^2(T) \sim T^{-1}$ decay.

We show a schematic representation of the behavior of the variance $\sigma^2(T)$ versus time in log log scale in Fig. 2.4.

### 2.2.3 Numerical Results

#### 2.2.3.1 Short-range network

In the SRN case we fix the nonlinearity strength $g = 1$ without loss of generality. The system size $N = 10^4$. At $\theta = 0$ the LOs are frozen – the system is at the integrable limit. Upon increasing the parameter $\theta$, the LOs get intercoupled with nearest neighbors into a SRN. We study the statistics of local and extended observables in the SRN in close proximity to the corresponding integrable limit. The variance $\sigma^2(T)$ of the set of time averages of EOs (a) and LOs (b) in the SRN for different $\theta$ is plotted in Fig. 2.5.

**Extended observables:** *Fig.2.5(a) Extended observables EOs with $k = \pi/2$. Three cases with $\theta = 0.1, 0.01, 0.001$ are shown and are practically not distinguishable. The dashed line indicates a $T^{-1}$ decay. The inset shows the local derivatives of all three curves.* All curves show the same result - an immediate decay $\sigma^2(T) \sim T^{-1}$ from the shortest averaging times on. This holds even in the integrable limit itself for $\theta = 0$. Similar results have been shown by [55] – due to the central limit theorem a random transformation of the set of action variables shows statistical properties similar to those of a mixing system when indeed it is not. Such a measurement does not necessarily imply true ergodicity, as follows from what comes next. It also does not allow for a measurement of the large but finite ergodization timescale $T_E$. The observed thermalization is in good agreement with the predictions in Fig. 2.3(a).

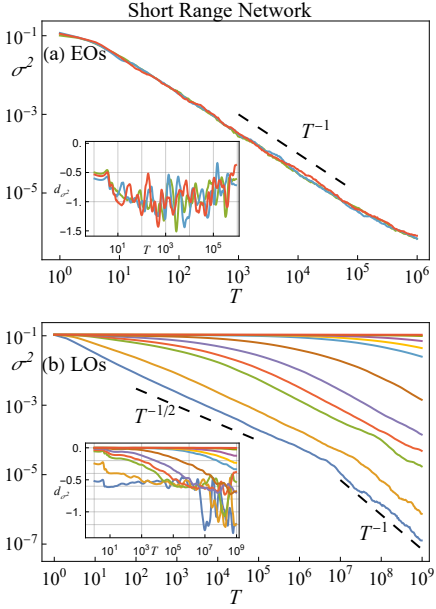
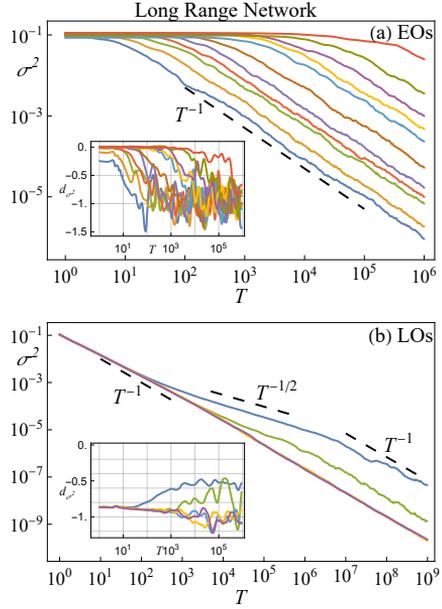

Figure 2.5: Numerical results in SRN case. See italic text for the details.

Figure 2.6: Numerical results in LRN case. See italic text for the details.

**Local observables:** *Fig.2.5(b) Local observables LOs and additional averaging over 10 trajectories (realizations). The parameter $\theta$ takes values {0.5, 0.25, 0.1, 0.075, 0.05, 0.025, 0.01, 0.0075, 0.005, 0.0025, 0.001} from bottom blue to top red. The two dashed lines indicate $T^{-1/2}$ and $T^{-1}$ decay respectively. The transi-* tion from $T^{-1/2}$ to $T^{-1}$ decay is observed for $\theta = 0.5, 0.25, 0.1$ *(blue, orange and green curves respectively). The inset shows the local derivatives of all curves.* At times $T < T_E$ we see approximately constant behavior $\sigma^2(T) \approx \sigma^2(0)$. For $T > T_E$ we observe the diffusive decay $\sigma(T) \sim T^{-1/2}$. For even larger averaging times $T > T_D$ we observe a transition to $\sigma^2(T) \sim T^{-1}$ due to finite size effects [51]. Clearly the ergodization time $T_E$ grows upon approaching the integrable limit. Together with the fast thermalization of EOs we arrive at a very good agreement of our observations with the prediction in Fig. 2.3(b).

### 2.2.3.2 Long-range network

In the LRN case we fix $\theta = 0.33\pi$ without loss of generality. The system size $N = 10^5$. For $g = 0$ the normal mode coefficients $|c_k|$ are constant in time. The EOs are frozen, indicating another integrable limit. Upon increasing $g \neq 0$ the EOs get intercoupled with an all-to-all coupling into a LRN. We study the statistics of local and extended observables in the LRN in close proximity to the corresponding integrable limit. The variance $\sigma^2(T)$ of the set of time averages of EOs (a) and LOs (b) in the LRN for different $g$ is plotted in Fig. 2.6.

**Extended observables:** *Fig.2.6(a) Extended observables EOs with $k = \pi/2$. The parameter $g$ takes values {0.5, 0.25, 0.1, 0.075, 0.05, 0.025, 0.01, 0.0075, 0.005, 0.0025, 0.001} from bottom blue to top red. The dashed line indicates a $T^{-1}$ decay. The inset shows the local derivatives of all curves.* At times $T < T_E$ we see approximately constant behavior $\sigma^2(T) \approx \sigma^2(0)$. For times $T > T_E$ we observe a $\sigma^2(T) \sim T^{-1}$ decay. Clearly the ergodization time $T_E$ grows upon approaching the integrable limit $g \to 0$. We arrive at a very good agreement of our observations with the prediction in Fig. 2.4(a).

**Local observables:** *Fig. 2.6(b) Local observables LOs and additional averaging over 10 trajectories (realizations). Four cases with $g = 0.5, 0.1, 0.005\ 0.0025$. Three dashed lines indicate $T^{-1}$, $T^{-1/2}$ and again $T^{-1}$ decay respectively. The transition from $T^{-1}$ to $T^{-1/2}$ and back to $T^{-1}$ decay is observed for $g = 0.5, 0.1$ (blue and green curves respectively). The inset shows the local derivatives of all curves.* All curves show the same result - an immediate decay $\sigma^2(T) \sim T^{-1}$ from the shortest averaging times on. This holds even in the integrable limit itself for

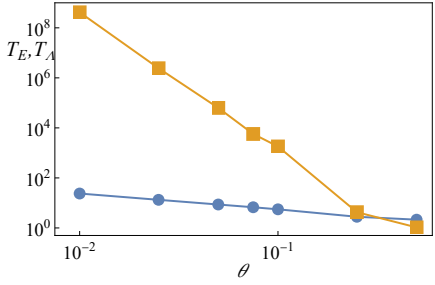
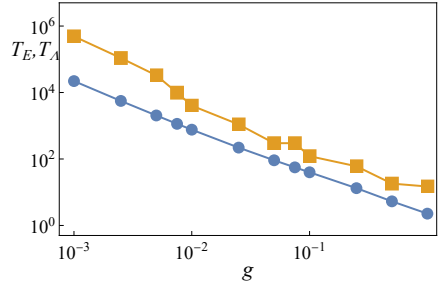

Figure 2.7: $T_E$ (blue circles) and $T_\Lambda$ in SRN case.

Figure 2.8: $T_E$ (blue circles) and $T_\Lambda$ in LRN case.

$g = 0$. Similar results have been shown by [55] – due to the central limit theorem a random transformation of the set of action variables shows statistical properties similar to those of a mixing system when indeed it is not. Such a measurement does not imply true ergodicity, as follows from what comes next. We also notice a transition to $\sigma^2(T) \sim T^{-1/2}$ starting at roughly $T \approx T_E$ at which EOs start to thermalize, therefore inducing diffusion in real space. A subsequent transition to $\sigma^2(T) \sim T^{-1}$ happens for $T > T_D$ due to finite size effects. The observed thermalization is in good agreement with the prediction in Fig.2.4(b).

### 2.2.3.3 Time scales

In both SRN and LRN cases we compare the ergodization times $T_E$ with the characteristic Lyapunov time scales $T_\Lambda$ of chaotic dynamics. As the variance of finite time averages of actions $\sigma^2(T)$ is expected to transition from nearly constant behavior to $\sigma^2(T) \sim T^{-1/2}$ in SRN and $\sigma^2(T) \sim T^{-1}$ in LRN we follow the local derivatives $d_{\sigma^2}$ of the variance curves $\sigma^2(T)$, see insets of Fig. 2.5 and Fig. 2.6. The ergodization time $T_E$ is extracted as the time when $d_{\sigma^2} = -0.25$ in SRN and $d_{\sigma^2} = -0.75$ in LRN for the first time. The Lyapunov times $T_\Lambda$ are computed for the same parameter values and plotted against $T_E$ in Fig. 2.7 for SRN and Fig. 2.8 for LRN. In the SRN (*Lyapunov time $T_\Lambda$ - blue circles and ergodization time $T_E$ - orange rectangles*) we notice $T_E \gg T_\Lambda$ such that the ratio $T_E/T_\Lambda$ grows quickly as the system approaches integrable limit $\theta \to 0$.

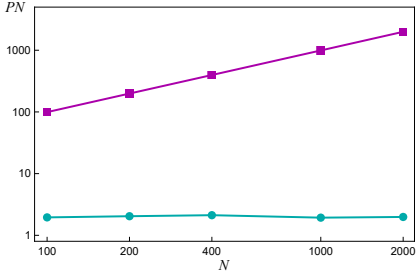 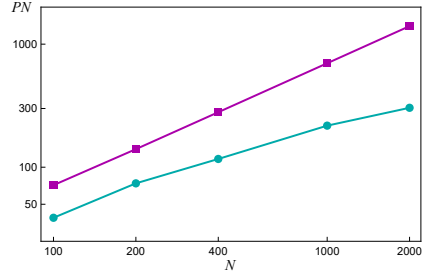

Figure 2.9: Participation number of tangent vector against system size in SRN. LOs representation: cyan circles; EOs representation: purple squares.

Figure 2.10: Participation number of tangent vector against system size in LRN. LOs representation: cyan circles; EOs representation: purple squares.

In the LRN (*Lyapunov time $T_\Lambda$ - blue circles and ergodization time $T_E$ - orange rectangles*) case $T_\Lambda \sim T_E$ and their ratio is practically constant upon approaching the integrable limit $g \to 0$.

#### 2.2.3.4 Lyapunov time and tangent vector

In Fig. 2.9 and Fig. 2.10 we plot the participation $PN$ number of the tangent vector corresponding to the largest Lyapunov exponent as a function of system size. For the SRN case the parameters are $g = 1$, $\theta = 0.001$. For the LRN case the parameters are $\theta = 0.33\pi$, $g = 0.001$. The participation numbers depend on the basis choice of the tangent vector. Thus we compute the $PN$ in both LO and EO representations.

In the SRN case (inset in Fig. 2.9) the tangent vectors in LO representation are localized due to rare chaotic resonances in real space. Therefore the $PN$ number in LO representation is expected to be system size independent. Since a localized distribution in real space turns delocalized in reciprocal space, the EO representation results in a $PN$ which scales with the system size.

In the LRN case ( inset in Fig. 2.10) resonances are expected to appear almost everywhere in the EO representation. Thus we expect the $PN$ number in EO representation to scale with the system size. This is in accord with our numerical observation. Interestingly the delocalized nature of the tangent vector in the EO representation does not imply that the LO representation will result in localized structures. Indeed the numerical computation show that the $PN$ number in LO representation also increases with the system size.

### 2.2.4 Preliminary Discussion

Our main findings from the observable dynamics is that the different choices of observables for a thermalization study can lead to ambiguous conclusions. In particular, we consider macroscopic systems which are tuned close to an integrable limit. Their slow adiabatic invariants are given by the conserved actions at the very integrable limit. Making these actions the observables of choice will obviously result in the correct thermalization analysis as argued in the Sec. 1.4.4. We study two different classes of weakly nonintegrable lattice systems with discrete translational invariance, in which the actions are interacting through either a long-range network, or a short-range network. For LRNs the actions are extended in real space, so the relevant observables are extended. For SRNs the actions are local in real space, thus the relevant observables are local. As we show in particular, the choice of the 'wrong' observables - LOs for LRNs or EOs for SRNs - results in a seemingly quick thermalization without any slowing down upon approaching the integrable limit, including the limit itself. At the same time the correctly chosen observables - EOs for LRNs and LOs for SRNs - shows a dramatic slowing down of thermalization. We also demonstrate that a simultaneous study of both LO and EO thermalization allows to unambiguously identify the slowing down, and even conclude which class - LRN or SRN - is under study.

As a result of our study, we extract the ergodization timescale $T_E$ as a function of the control parameter which tunes the distance from the integrable limit. We also compute the largest Lyapunov exponent and its inverse - the Lyapunov time $T_\Lambda$. It follows that the LRN class is characterized by only one timescale as $T_E \sim T_\Lambda$. At variance to the above, the SRN class must be characterized by other diverging scales, as we observe that $T_E \gg T_\Lambda$ in accord with previous observations for Josephson junction arrays [49] and Klein-Gordon chains [50]. This second time or lengthscale was predicted to arise from low densities of rare resonances in real space, and the need for these resonances to migrate over the increasing and diverging average distance between them [49], see Sec. 1.4.3.

For times larger than the ergodization timescale $T_E$ the finite time average distributions of LOs show a diffusive convergence of their variance $\sigma^2 \sim T^{-1/2}$. The impact of a finite size $N$ of the system results in a diffusion timescale $T_D \sim N^2$ when the variance crosses over from to $\sigma^2 \sim T^{-1}$. These findings are in line with studies on Hamiltonian dynamics such as Josephson junctions [51].

The above scheme of identifying the correct network class relies on measuring both LOs and EOs and on varying the control parameter of the distance to the integrable limit. Interestingly we can tell the right network class also if we do not vary that control parameter, but instead vary the system size $N$. For that we note that the computation of the largest Lyapunov exponent comes with its corresponding tangent vector information. We use this information to compute its average participation number $PN$ in both the direct local space, and in reciprocal space. For the SRN we already expect the tangent vector to be highly local in real space, thus delocalized in reciprocal space. Indeed, $PN$ is essentially independent of $N$ in direct space, but scales $PN \sim N$ in reciprocal space. At variance to that, the LRN results in a $PN \sim N$ scaling for both spaces. Therefore we can tell the network class from a finite size analysis of the participation number of the tangent vector, without varying the distance to the integrable limit.

## 2.3 Lyapunov spectrum scaling for classical many-body dynamics close to integrability

There are serious limitations to studying thermalization through observable dynamics. As discussed above the choice of observables might be ambiguous [48, 68], and even for integrable systems specifically chosen observables show ergodic thermal-like behavior [28]. The analysis performed in the previous section allows for identification and classification of weakly nonintegrable dynamics. However it relies on the preexisting knowledge of the actions of the integrable limit and the necessity to perform full scale analysis of both local and extended observables. In this section we propose an alternative viewpoint on classification of weakly nonintegrable dynamics by studying the properties of the Lyapunov spectra in proximity to integrable limits. We show that the scaling properties of the Lyapunov spectrum offer a conceptually novel way for the description of weakly nonintegrable dynamics in a generic model setup. We consider a macroscopically large system and characterize thermalization in both SRN and LRN regimes, thus drawing a very general picture that encapsulates a great number of physically realizable scenarios and is directly applicable to most weakly nonintegrable classical systems.

Resolving the entire Lyapunov spectrum for a large system is a numerically challenging task. As is discussed in a greater detail below, it relies on the simultaneous evolution of a large number of trajectories which scales with the system size, [54]. In addition as observed on multiple occasions the thermalization dynamics in general slow down close to integrability, and may even cease to be observed [91, 92, 93, 94, 95, 96, 97], which was also noted in earlier studies of dynamical systems [17, 33, 66, 98]. This adds up to the computational complexity of the task. Here the unitary circuit maps discussed above come in as a great asset for the fast, exact and error-free discrete-time evolution. We consider a system of $N = 200$ sites and compute the full LCE spectrum in both SRN and LRN limits. It turns out that the SR and LR networks are characterized by qualitatively distinct spectra which is discussed below in detail. Below we present our findings based on Ref. [99].

### 2.3.1 Methods

#### 2.3.1.1 Lyapunov spectrum

In Sec. 1.3.2 of the introduction we discussed some basic ideas behind exponential divergence of nearby trajectories and their characterization using the Lyapunov exponent. Practically this has been done by means of extracting the exponent of growth over time of a deviation vector $\mathbf{w}$. However this approach is only a first step of a far more general framework of the Lyapunov spectra. One could broaden the perspective and characterize the *change of phase space volume* of dimension $p \leq 2N$ along a reference trajectory $\mathbf{x}(t)$. For this one introduces a *set* of linearly independent normal deviation vectors $\{\mathbf{w}_n\}_{n=1}^p$. Then the volume is given by a cross product of all vectors $V_p = ||\mathbf{w}_1 \times \mathbf{w}_2 \times \mathbf{w}_3 ... \times \mathbf{w}_p||$. And similarly to the definition of the rate of growth of a vector $r(t) = ||\mathbf{w}(t)||/||\mathbf{w}(0)||$ we define a rate of growth of the volume $r_p(t) = V_p(t)/V_p(0)$ and the corresponding **Lyapunov characteristic exponent of order** $p$, alternatively known as $p-$LCE:

$$\lim_{t \to \infty} \frac{1}{t} \log r_p(t) \equiv \Lambda^{(p)} \qquad (2.24)$$

It is clear now that the exponent defined for one deviation vector $\mathbf{w}$ is nothing but a 1–LCE discussed before, which in this context can be denoted as $\Lambda^{(1)}$ – a specific case for a more general definition in Eq. (2.24).

In 1968 Osceledec proved his multiplicative ergodic theorem [100] which turned out to be of a great importance for the computation of Lyapunov exponents. An important consequence of this theorem comes in terms of the following equality:

$$\Lambda^{(p)} = \sum_n^p \Lambda_n^{(1)}, \qquad (2.25)$$

where $\Lambda_n^{(1)}$ is a 1–LCE corresponding to the vector $\mathbf{w}_n$ from a set of linearly independent vectors $\{\mathbf{w}_n\}$. If $p = 2N$, i.e the whole the phase space is considered then the set of $1-$LCEs $\{\Lambda_n^{(1)}\}_{n=1}^{2N}$ forms the **Lyapunov spectrum**. Due to the fact that any $\Lambda^{(p)}$ can be represented as a sum of $\Lambda_n^{(1)}$ usually the 1–LCE are discussed as characteristic exponents and are referred to as just *LCEs* and are written without the superscript as $\Lambda_n^{(1)} \equiv \Lambda_n$. We also follow this notation. Without proof (but we refer readers to [52]) we point out that Lyapunov exponents come ordered: $\Lambda_1 \geq \Lambda_2 \geq \Lambda_3 \geq ... \geq \Lambda_{2N}$. And due to this ordering $\Lambda_1 \equiv \Lambda_{max}$ coinciding with $\Lambda^{(1)}$ is called the **largest Lyapunov exponent**.

The numerical approach for computing the entire Lyapunov spectrum was proposed by Benettin [54] using an orthonormal set of vectors $\{\mathbf{w}_n(t)\}$ and extraction of the growth coefficients $\{r_n(t)\}$. As opposed to the evolution of a single trajectory now $2N+1$ trajectories have to be propagated in time. This significantly increases the computational complexity. Additionally the algorithm requires application of the orthonormalization Gram-Schmidt procedure which further increases computation time.

#### 2.3.1.2 Properties of Lyapunov spectra

Immediately we notice the following property of the Lyapunov spectra for the phase space volume preserving dynamics:

$$\sum_n \Lambda_n = 0. \tag{2.26}$$

This is due to the corresponding $\Lambda^{(2N)} = 0$ in Eq. (2.24).

In Ref. [54] Benettin et al. provided a proof for the symmetry of the spectrum:

$$\Lambda_n = -\Lambda_{2N-n+1}. \tag{2.27}$$

Another important property of Lyapunov exponents relates them to the conservation laws.

$$\Lambda_n = -\Lambda_{2N-n+1} = 0 \quad - corresponds\ to\ a\ conservation\ law. \tag{2.28}$$

This is due to the fact that deviation vectors $\mathbf{w}$ in the directions perpendicular to the hypersurfaces arising from the conservation laws results in the absence of exponential growth, which is characterized by vanishing Lyapunov exponents.

The positive part of Lyapunov exponents is related to the measure of chaoticity the Kolmogorov-Sinai entropy [6]:

$$K_{\text{KS}} = \sum_{\Lambda_n > 0} \Lambda_n. \qquad (2.29)$$

The great advantage in using the LCE spectra as a tool in characterizing chaotic dynamics is their independence from the choice of observables and thus avoiding the ambiguity discussed in the above sections. The disadvantage as discussed comes with the complexity of numerical computation of the entire spectrum. Below we elaborate on the utility of the LCE spectra and provide several examples for their usage as a diagnostic tool for dynamical properties of physical systems.

### 2.3.1.3 Utility of Lyapunov spectra

As discussed, for detecting the chaotic dynamics it is sufficient to evaluate only one exponent $\Lambda_{max}$. The utility of the entire Lyapunov spectra comes as a tool to identify conservation laws. Moreover, due to the property given in Eq. (2.28) one is able to identify exactly the number of conserved quantities. To illustrate this we show a figure from Ref. [101]. The authors consider the discrete nonlinear Schrödinger equation which is characterized by two integrals of motion: the total energy and the total norm. Computing the Lyapunov spectrum results in corresponding four vanishing Lyapunov exponents (see cyan line in Fig. 2.11). Then a breather-like excitation is added in the middle of the chain effectively creating two subsystems with their own energy and norm conserved. The energy of the breather is conserved as well. In total one expects ten vanishing LCEs instead of the initial four. And indeed this indicated in the observed Lyapunov spectrum (see. Fig. 2.11).

### 2.3.1.4 Evolution

We consider the unitary map model with nonlinearity (see Fig. 2.1). The evolution of a reference trajectory is performed by Eq. (2.5). We use periodic boundary conditions $\psi_{N+1} = \psi_1$. The initial conditions for the amplitudes of the local complex components are drawn from an exponential distribution $p(x) = xe^{-x^2}$, while their phases were generated as uncorrelated and random numbers chosen uniformly from the interval $[0, 2\pi]$. The state vector is then uniformly rescaled such that the norm density $\frac{1}{N}\sum|\psi_n|^2 = 1$. The largest integration time varied between $t_{\max} = 10^8$ and $t_{\max} = 10^9$. We have performed computations for a set of initial conditions to ensure the independence of results on the choice of initial state. Unless stated otherwise the system size is set to $N = 200$.

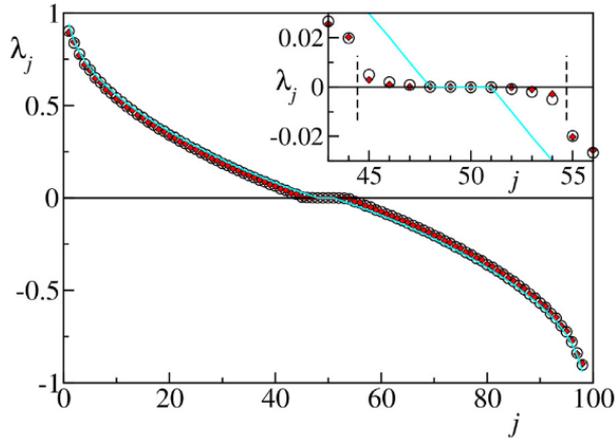

Figure 2.11: The Lyapunov spectrum for discrete nonlinear Schrödinger equation taken from [101]. The inset shows zoomed-in central part of the spectrum. Cyan line indicates a clean system with 2 conserved quantities and respectively 4 vanishing LCEs. Upon adding a "defect" breather 6 more vanishing LCEs appear.

### 2.3.1.5 Lyapunov spectrum computation

We compute the full LCE spectra using methods described in [54, 52] in both SRN and LRN limits upon variation of $\theta$ and $g$ respectively. We introduce a set of deviation vectors $\{\mathbf{w}_i\}$ and evolve them in the corresponding phase space according to the tangent equations derived in Sec. 2.1.4. We extract the Lyapunov exponents corresponding to their growth $r_i(t) = ||\mathbf{w}_i(t)||$. For each vector we compute the transient value $X_i(t) = 1/t \sum_\tau^t \log r_i(\tau)$ which in the infinite time limit turns into the Lyapunov characteristic exponent (LCE) $\Lambda_i = \lim_{t\to\infty} X_i(t)$. The numerically computed LCEs are the values of $X(t)$ extracted at the last step of the dynamics. In Fig. 2.12 we show the evolution and saturation of the transient values $X_i(t)$ against time for the positive part of the Lyapunov spectrum in both SRN and LRN cases. This is similar to Fig. 1.3 given in the introductory Sec. 1.3.2. Due to the conservation of the norm two exponents are expected to attain zero value. In the figure we see one of them (bottom most purple line) tending to zero with increasing time and no saturation.

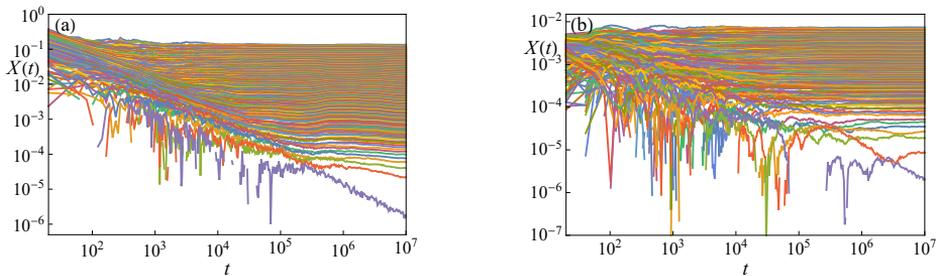

Figure 2.12: The evolution of positive transient Lyapunov exponents. a) SRN case with angle $\theta = 0.1$ and nonlinearity $g = 1.0$, b) LRN case with angle $\theta = 0.33\pi$ and $g = 0.1$. For both cases system size $N = 200$.

The number of Lyapunov characteristic exponents equals the dimensionality of the phase space, in our case $4N$. Without loss of generality we consider only positive LCEs. We renormalize Lyapunov spectra $\overline{\Lambda}_i = \Lambda_i/\Lambda_{max}$ and rescale the index index $\rho = i/2N$ so that all positive LCEs $\overline{\Lambda}(\rho)$ correspond to $\rho \in [0,1]$.

### 2.3.2 Results

First, we show the dependence of the largest Lyapunov exponent $\Lambda_{max}$ on the distance $g$ or $\theta$ to the integrable limit in Fig. 2.13 for both networks. Both curves show a dependence which might resemble a power law $\Lambda_{max} \sim g^\nu$ and $\Lambda_{max} \sim \theta^\mu$ with $\nu \approx 1/2$ and $\mu \approx 3/2$. Remarkably the SRN case shows a much slower diminishing of $\Lambda_{max}$ upon approaching the integrable limit as compared to the LRN. This is similar to the study of a Hamiltonian system dynamics [50, 102, 103]. Our data in Fig.2.13 are obtained for two different system sizes $N = 100, 200$ and show very good agreement, therefore we can exclude finite-size corrections.

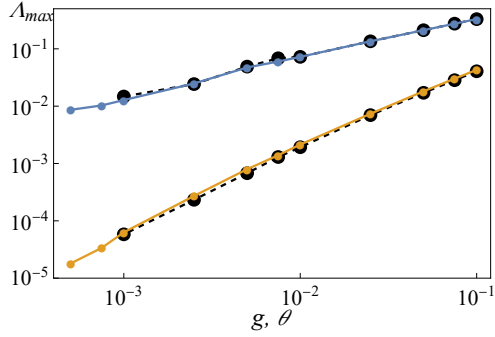

Figure 2.13: The largest Lyapunov exponents $\Lambda_{\text{max}}$ in SRN (blue small circles, top) and LRN (orange small circles, bottom) regime versus the corresponding deviation from integrable limit $g, \theta$. Solid lines connect the data points and guide the eye. For the SRN case, the parameter nonlinearity is fixed $g = 1.0$, while for the LRN case the angle $\theta$ is fixed at $0.33\pi$. For both cases system size $N = 200$. The large black circles connected by dashed lines correspond to the data for system size $N = 100$.

We now proceed to the analysis of the entire Lyapunov spectrum. In Fig. 2.14 and Fig. 2.16 we show the renormalized Lyapunov spectrum $\overline{\Lambda}_i = \Lambda_i/\Lambda_{\text{max}}$ for SRN and LRN respectively. We notice a dramatic qualitative difference between the two regimes. For the LRN case the renormalized Lyapunov spectrum $\bar{\Lambda}(\rho)$ converges to a limiting smooth curve for $g \to 0$. For the SRN instead that curve vanishes in an exponential way. We explain these observations in detail below.

For the SRN an increasing number of Lyapunov exponents seems to be vanishing upon approaching the integrable limit as seen in Fig. 2.14. *The nonlinearity strength is fixed at $g = 1.0$. Angle $\theta$ varies from $10^{-1}$ (blue, top) to $5 \cdot 10^{-4}$ (red, bottom). The inset shows the coefficient $1/\beta$ of the exponential decay (see* also Fig. 2.15) of the curves as a function of $\theta$. System size $N = 200$. We replot the same spectrum in log scale in Fig. 2.15 and notice an exponential decay of the renormalized spectrum:

$$\Lambda_\rho^{\text{SRN}} = \Lambda_{\max} e^{-\rho/\beta}. \tag{2.30}$$

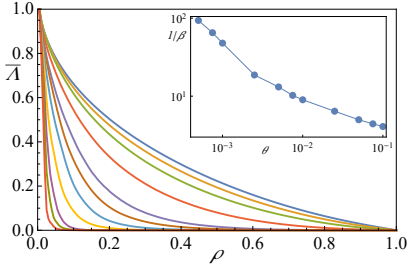 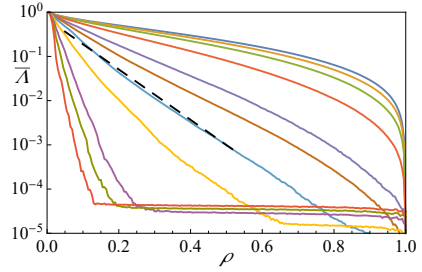

Figure 2.14: Renormalized Lyapunov spectra for SRN case. The inset shows the coefficient of the exponential decay, see Fig. 2.15.

Figure 2.15: Renormalized Lyapunov spectra for SRN case in a log scale corresponding to the data in Fig. 2.14.

We fit the exponential decay and plot the exponent $1/\beta$ versus $\theta$ in the inset in Fig. 2.14. We observe that the exponent is rapidly diverging upon approaching the integrable limit such that $\beta(\theta \to 0) \to 0$. The entire Lyapunov spectrum of the SRN is therefore characterized by two scaling parameters - the largest Lyapunov exponent $\Lambda_{max}$ which is an inverse timescale, and the parameter $\beta$ which is an inverse lengthscale. This result explains and agrees with previous studies on dynamical glass in Hamiltonian systems [50, 49] where the largest Lyapunov exponent stems from local resonances with rapidly increasing distance between them upon approaching the integrable limit. Our results show that the Lyapunov spectrum contains the quantitative scaling parameters of that dynamical glass theory.

In contrast, the LRN spectrum is characterized by a single parameter scaling. The renormalized Lyapunov spectrum approaches a smooth limiting curve $\bar{\Lambda}(\rho)$ as seen in Fig. 2.16. We compute the limiting curves by a linear fit of $\bar{\Lambda}_\rho(g)$ at

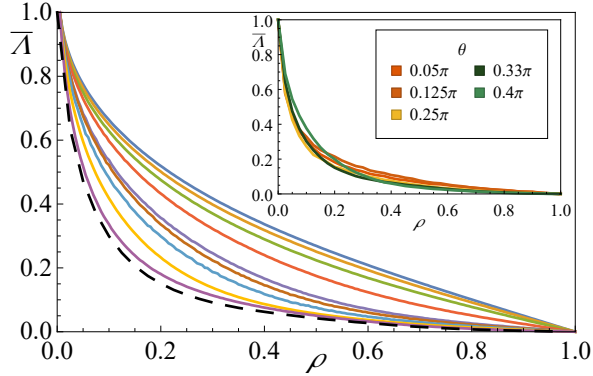

Figure 2.16: Renormalized Lyapunov spectrum $\{\bar{\Lambda}_i\}$ for LRN in proximity to the integrable limit. The angle $\theta = 0.33\pi$ is fixed. The deviation from integrable limit $g$ varies from $10^{-1}$ (blue) to $10^{-3}$ (purple). The dashed line is to show the asymptotic curve as $g \to 0$. In the inset we showcase the asymptotic curve as parameter $\theta$ is varied. For all cases system size $N = 200$.

each value of $\rho_j$. Thus in the LRN regime, the final form of the spectrum is given by:

$$\Lambda_\rho^{\text{LRN}} = \Lambda_{\max} f(\rho, \theta) \tag{2.31}$$

The limiting curves for different values of $\theta$ are plotted in the inset of Fig. 2.16 and show little if any variation. It appears that the limiting curve $f(\rho)$ is universal for all LRN parameter choices.

To further characterize the chaotic dynamics and showcase the difference between SR and LR networks we compute the Kolmogorov-Sinai entropy $K_{\text{KS}} = \int_0^1 \bar{\Lambda}_\rho d\rho$. In the SRN case, from Eq. (2.30) follows

$$K_{\text{KS}}^{\text{SRN}} = \Lambda_{\max} \beta (1 - e^{-1/\beta}) . \tag{2.32}$$

Therefore the renormalized Kolmogorov-Sinai entropy $k_{\text{KS}} = K_{\text{KS}}/\Lambda_{max}$ tends to zero in the integrable limit $k_{\text{KS}} \approx \beta$.

In the LRN regime the integral over the asymptotic function $f(\rho, \theta)$ (see Eq.

([2.31](#))) leads to finite values of the renormalized KS entropy $k_{\text{KS}} = \int_0^1 f(\rho)d\rho > 0$ at the very integrable limit.

### 2.3.3 Discussion

In addition to the standard methods of analysis of equilibration of macroscopic systems we identified the Lyapunov spectrum as a universal characteristic descriptor of the complex phase space dynamics of a macroscopic system in proximity to an integrable limit. Such an analysis offers independence from the choice of observables and allows for identification of mixing dynamics uniquely. As for the classes of integrability, the long-range networks are characterized by a single parameter scaling of the Lyapunov spectrum - knowing the largest Lyapunov exponent allows to reconstruct the entire spectrum. Consequently all Lyapunov exponents scale as the largest one upon approaching the integrable limit. As discussed typical long-range networks are realized with translationally invariant lattice systems in the limit of weak nonlinearity. Thus we expect similar results for Hamiltonian systems with weak nonlinearity. In that case, the actions correspond to normal modes extended over the entire real space. Nonintegrable perturbations will typically couple them all. On the other side, short-range networks are characterized by a two-parameter scaling. In addition to the largest Lyapunov exponent, a diverging lengthscale results in a suppression of the renormalized Lyapunov spectrum upon approaching the integrable limit.

## 2.4 Other integrable limits and outlook

Up to now we have discussed two integrable limits stemming from the nature of nonintegrable perturbation to otherwise integrable, nonchaotic dynamics. The long-range networks originate from nonlinear perturbations coupling the normal modes of the system. While the short-range network come from finite range of the coupling between local coordinates. These two classes cover a broad range of dynamical models. The FPUT [18] and the FPUT like scenarios belong to LRN, while any translationally invariant lattice in the limit of small coupling between the sites belongs to SRN. The chain of thought continues with a reasonable question: do there exist other integrable limits? Do all models fall into classes mentioned above? Below we showcase some scenarios to give an outlook for the answer to this questions. First, we show a peculiar regime of free propagation which can be called 'free gas' integrable limit. This regime is not yet understood in the sense of belonging to SRN or LRN integrability classes. Next we put the understanding of SR and LR networks to the test in the disordered dynamics perturbed by nonlinearity. Finally we apply the approach developed so far to Hamiltonian systems and report on the universality of Lyapunov spectrum computation approach.

### 2.4.1 Free propagation integrable limit

First we cover the so called free propagation integrable limit found in the Unitary Circuits map. We recall the equations of motion (2.5) for the map below:

$$\psi_n^A(t+1) = e^{ig|\varphi_n^A(t)|^2}\varphi_n^A(t) \quad , \quad \psi_n^B(t+1) = e^{ig|\varphi_n^B(t)|^2}\varphi_n^B(t) \quad (2.33)$$

where $\varphi_n^{A,B}(t)$ are the components of state vector $\vec{\Psi}$ after the application of mixing maps $\hat{C}$:

$$\varphi_n^A(t) = \left[\cos^2\theta\psi_n^A(t) - \cos\theta\sin\theta\psi_{n-1}^B(t) + \sin^2\theta\psi_{n+1}^A(t) + \cos\theta\sin\theta\psi_n^B(t)\right]$$

$$\varphi_n^B(t) = \left[\sin^2\theta\psi_{n-1}^B(t) - \cos\theta\sin\theta\psi_n^A(t) + \cos^2\theta\psi_n^B(t) + \cos\theta\sin\theta\psi_{n+1}^A(t)\right]$$
(2.34)

In the limit $\theta = \pi/2$ the $A$ and $B$ sublattices of the system are decoupled:

$$\psi_n^A(t+1) = e^{ig|\psi_{n+1}^A(t)|^2} \psi_{n+1}^A(t)$$
$$\psi_n^B(t+1) = e^{ig|\psi_{n-1}^B(t)|^2} \psi_{n-1}^B(t)$$
(2.35)

The components of each sublattice are translated unchanged through the system without interaction with the other sublattice. This is similar to the free propagation of the gas particles, hence the term 'free gas' limit. The individual components of the sublattice are also decoupled. In the co-moving frame of sublattice the norm of each component of sublattice is conserved over time even in the presence of nonlinearity. From the standpoint of the dispersion relation Eq. (2.7) this limit corresponds to the linear Dirac spectrum. Deviation from such a limit $\pi/2 - \theta \equiv \varepsilon \ll 1$ corresponds to the coupled equations:

$$\psi_n^A(t+1) = e^{ig|\varphi_n^A(t)|^2} \varphi_n^A(t),$$
$$\varphi_n^A(t) = \left[\psi_{n+1}^A(t) - \varepsilon(\psi_{n-1}^B(t) - \psi_n^B(t))\right],$$

$$\psi_n^B(t+1) = e^{ig|\varphi_n^B(t)|^2} \varphi_n^B(t),$$
$$\varphi_n^B(t) = \left[\psi_{n-1}^B(t) + \varepsilon(\psi_{n+1}^A(t) - \psi_n^A(t))\right].$$
(2.36)

The equations above are similar to those of a short-range limit network given in Eq. (2.12). However at variance to the SRN case the local norm is not conserved in this limit. As discussed in the previous sections addressing the weakly nonintegrable network starts with identification of the actions in the integrable limit. For this case we are yet to find the transformation to bring Eqs. (2.36) to the form of conserved actions and identify the corresponding coupling network.

Here the utility of the Lyapunov spectra as a tool for the study of weakly nonintegrable dynamics comes with a great advantage: one does not necessarily need to know the exact form of the actions of the integrable limit to study the chaotic properties of the system.

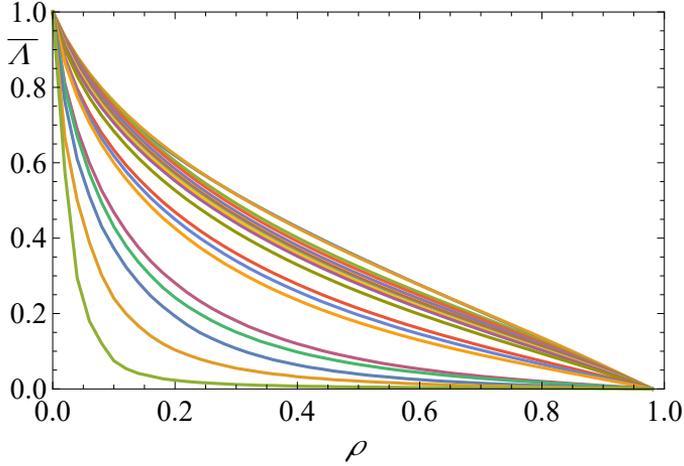

Figure 2.17: Renormalized Lyapunov spectrum $\{\bar{\Lambda}\}$ against the rescaled index $\rho$ in proximity to the 'free gas' integrable limit. The nonlinearity strength is fixed at $g = 1.0$. The deviation parameter $\varepsilon$ varies from $0.5$ (orange, top) to $10^{-5}$ (green, bottom). The system size is fixed at $N = 50$.

In Fig. 2.17 we show the Lyapunov spectra for the system weakly deviated from the free propagation regime. We were not able to identify the universal curve corresponding to the LRN case. At the same time the exponential decay with almost vanishing Lyapunov exponents which is characteristic for the SRN case is also not present. Thus far the system in proximity to the 'free propagation' limit seems to belong to its own class of networks and has to be studied in more detail.

### 2.4.2 Disordered systems with nonlinearity

#### 2.4.2.1 Anderson localization

In 1958 Philip Anderson studied the transport of a single particle in disordered medium [104]. The goal was to construct a minimal model adequately simulating the disorder and irregularity typical for most macroscopic materials. Such a model comes in terms of a tight-binding chain with random onsite potential:

$$H = \sum_n \varepsilon_n c_n^\dagger c_n - t \sum_n c_n^\dagger c_{n+1} + \text{h.c.}, \tag{2.37}$$

where $c_n$ are annihilation operators. The disorder is encoded in random site-dependent values of local energy $\varepsilon_n$ that are drawn independently and uniformly from $[-W, W]$. In literature $W$ is referred to as the strength of disorder. Anderson discovered that all eigenstates of Eq. (2.37) are exponentially localized, so at $n \to \infty$:

$$|\Psi_n| \sim e^{-|n|/\xi}, \tag{2.38}$$

where $\xi$ is called localization length and depends on the energy of the eigenstates. The discovery had great implications on the studies of transport phenomena and now referred to as the Anderson localization (AL).

Since the discovery of the phenomenon the focus shifted on the possible breakdown of localization. The natural extension of the model to multiple dimensions showed no localization in two dimensions [105]. The three-dimensional case on the other hand shows a transition to delocalized state for some critical value of disorder strength $W_c$ [106]. For an overview of Anderson Localization we recommend readers the Ref. [107].

#### 2.4.2.2 Weak nonintegrability networks

Introducing many-body interaction is a natural extension of the single particle disordered model discussed above. Despite expected delocalization recent observations show a persistent many-body localized (MBL) state [56, 57, 58]. Understanding and characterizing these finding presents a great challenge in modern physics.

We take an interest in disordered many-body dynamics from classical perspective, i.e. disordered nonlinear dynamics. Some extensive investigations have been performed on nonlinearity induced wave-packet spreading and transport [53, 108, 109]. However, there have been observations of highly resistive 'bad metal' regime in Josephson junction chains [59]. We suggest a possibility of highly resistive phase being nothing but SRN dynamics with extremely large thermalization times. Below we illustrate this idea using the aforementioned tight-binding model with addition of nonlinearity:

$$i\dot{\psi}_n = \varepsilon_n \psi_n - t(\psi_{n-1} - \psi_{n+1}) + g|\psi_n|^2 \psi_n. \tag{2.39}$$

As in the original model the onsite potential $\varepsilon_n$ is sampled from uniform distribution $[-W, W]$. The nonlinear term is introduced in line with mean-field philosophy. Recasting the equation in terms of the single particle Anderson eigenstates $\psi_n(t) = \sum_\nu c_\nu(t) \tilde{\psi}_n^\nu$, $\tilde{\psi}_n^\nu \sim e^{-|n_\nu - n|/\xi_\nu}$ ($n_\nu$ is a site around which the state with a number $\nu$ is localized) yields the following equations of motion for the coefficients of Anderson modes $c_\nu$:

$$i\dot{c}_\nu = E_\nu c_\nu + g \sum_{\nu_1,\nu_2,\nu_3} I_{\nu,\nu_1,\nu_2,\nu_3} c_{\nu_1}^* c_{\nu_2} c_{\nu_3}. \tag{2.40}$$

Here $E_\nu$ are the energies of the Anderson modes and the $I_{\nu,\nu_1,\nu_2,\nu_3}$ are the overlap integrals:

$$I_{\nu,\nu_1,\nu_2,\nu_3} = \sum_n \tilde{\psi}_n^\nu \tilde{\psi}_n^{\nu_1} \tilde{\psi}_n^{\nu_2} \tilde{\psi}_n^{\nu_3}. \tag{2.41}$$

Similarly to the translationally invariant systems the linear Anderson model with $g = 0$ behaves as integrable with conserved coefficients $c_\nu$ corresponding to the eigenstates of the system. Nonlinearity couples the Anderson modes causing delocalization [108]. The case of weak nonlinearity $g \ll 1$ may be viewed as a perturbation to the integrable dynamics. The equations above strongly resemble those of long-range network setups, see Eq. (1.28), with the difference coming in terms of the overlap integrals. In the LRN setups discussed so far the overlap terms come from conservation of momenta, see Eq. (1.29), and result in the absence of coupling between particular triplets of modes. However in the case of the disordered models the terms $I_{\nu,\nu_1,\nu_2,\nu_3}$ correspond to the overlap of exponentially decaying states:

$$I_{\nu,\nu_1,\nu_2,\nu_3} = \sum_n e^{-\frac{|n_\nu - n|}{\xi_\nu}} e^{-\frac{|n_{\nu_1} - n|}{\xi_{\nu_1}}} e^{-\frac{|n_{\nu_2} - n|}{\xi_{\nu_2}}} e^{-\frac{|n_{\nu_3} - n|}{\xi_{\nu_3}}}. \tag{2.42}$$

It is clear that if at least one of the triplet states, without loss of generality $\nu_1$, is localized far from the state $\nu$, i.e. $|n_\nu - n_{\nu_1}| > max(\xi_\nu, \xi_{\nu_1})$ then the overlap integral is vanishing. Thus the overlap integral couples a reference state $\nu$ with the *localization length dependent number of states*. This statement can be implemented in Eq. (2.40) as the cutoff in the number of the coupled terms:

$$i\dot{c}_\nu = E_\nu c_\nu + g \sum_{\nu_1,\nu_2,\nu_3}^{n_{\nu_{1,2,3}} \in (n_\nu \pm \xi_{max})} I_{\nu,\nu_1,\nu_2,\nu_3} c^*_{\nu_1} c_{\nu_2} c_{\nu_3}. \tag{2.43}$$

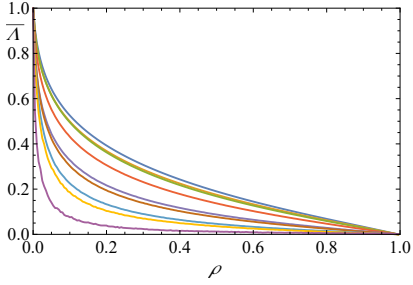 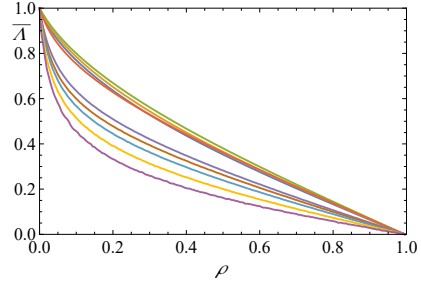

Figure 2.18: Renormalized Lyapunov spectrum in disordered case with $\xi \approx 3$.

Figure 2.19: Renormalized Lyapunov spectrum in disordered case with $\xi \approx 20$.

The cutoff is imposed by the largest localization length $\xi_{max}$. Here is the next step in our analysis: if localization length is of the order or larger then the system size $\xi_{max} > N$ then *all states are weakly intercoupled* which will result in the long-range network scenario. At variance in the case of localization length smaller then the system size $\xi_{max} < N$ *the number of coupled terms does not scale with the system size* $N$ falling under the definition of short-range networks.

To test the reasoning discussed above we use the unitary circuit map with non-linearity. In addition to the standard evolution given by Eq. (2.33) we introduce a disorder inducing map:

$$\hat{D}_n \psi_n^{A,B}(t) = e^{i\gamma_n^{A,B}} \psi_n^{A,B}(t), \tag{2.44}$$

where the local angles $\gamma_n^{A,B}$ are sampled from $[-W, W]$ in analogy to the random onsite potential discussed above. The unitary maps possess several advantages for studying the disordered systems. One is computational, unitary maps demonstrated record-breaking evolution times for nonlinear wave-packet spreading in disordered systems [88]. Another feature unique to unitary map systems is a regime of maximally disordered system with $W = \pi$. In this case all eigenstates have the same localization length $\xi$ which is uniquely determined from the mixing parameter $\theta$ [71, 110]:

$$\xi = -\frac{1}{\log|\sin\theta|}. \tag{2.45}$$

The unique localization length allows to consider all eigenstates as equivalent. Thus performing the analysis on statistics of observables as discussed in Sec. 1.4 is simplified compared to the typical energy dependent modes, such as in the translationally invariant dispersive systems or systems with disorder such as discrete nonlinear Schrödinger equation discussed above.

The equations of motion for the Anderson mode of unitary circuit map in the presence of nonlinearity are given as follows:

$$c_\nu(t+1) = e^{i\omega_\nu} c_\nu(t) + ig \sum_{\nu_1,\nu_2,\nu_3}^{n_{\nu_{1,2,3}} \in (n_\nu \pm \xi)} e^{i(\omega_{\nu_1}+\omega_{\nu_2}-\omega_{\nu_3})} I_{\nu,\nu_1,\nu_2,\nu_3} c_{\nu_1}(t) c_{\nu_2}(t) \left(c_{\nu_3}(t)\right)^*, \tag{2.46}$$

where $\omega_\nu$ are eigenfrequencies and $I_{\nu,\nu_1,\nu_2,\nu_3}$ overlap integrals:

$$I_{\nu,\nu_1,\nu_2,\nu_3} = \sum_n e^{-\frac{|n_\nu - n|}{\xi}} e^{-\frac{|n_{\nu_1} - n|}{\xi}} e^{-\frac{|n_{\nu_2} - n|}{\xi}} e^{-\frac{|n_{\nu_3} - n|}{\xi}}. \tag{2.47}$$

As discussed the overlap integrals corresponding to triplets where at least one state is localized outside the volume $n_\nu \pm \xi$ are considered negligible. In what follows we consider two regimes with $\xi \ll N$ and $\xi \sim N$ for varying nonlinearity strength $g$. In each case the limit $g = 0$ corresponds to integrability. Using Lyapunov spectra analysis we expect to detect an SRN-like behavior in the former and an LRN-like case in the latter.

In Fig. 2.18 and Fig. 2.19 we plot Lyapunov spectra for the system of size $N = 200$. The mixing parameter $\theta = 0.25\pi$ and $\theta = 0.4\pi$ corresponding to localization length $\xi \approx 3$ and $\xi \approx 20$ determined from Eq. (2.45) for disorder strength $W = \pi$. In both cases the parameter controlling the proximity to the integrable limit is the nonlinearity $g$ which assumes values in range $10^{-3} - 10^{-1}$ (bottom to top curves). We notice qualitatively different behavior of the spectra.

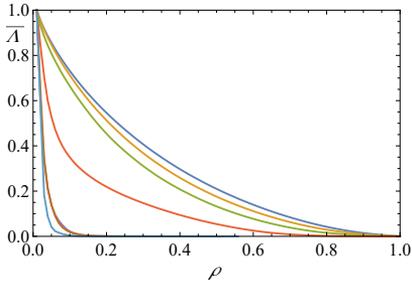 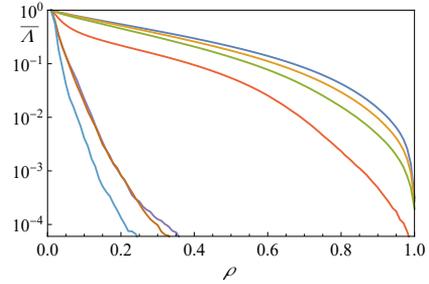

Figure 2.20: Hamiltonian dynamics. Renormalized Lyapunov spectra for SRN case. See Fig. 2.21 for a log scale.

Figure 2.21: Hamiltonian dynamics. Renormalized Lyapunov spectra for SRN case in a log scale corresponding to the data in Fig. 2.20.

The case of small localization length shows quick decay characteristic of the SRN networks (see Fig. 2.14), however the exponential decay with almost vanishing Lyapunov exponents is not observed. The large localization length case does not show quick decay and resembles the LRN results discussed above (see. Fig. 2.16).

Recently there has been a growing interest in studies of many-body systems with disorder and the many-body localization. Introducing nonlinearity serves as an approximation of the many-body interacting systems. Such cases can be analyzed using the framework we have developed. For large systems with some typical values of disorder the condition $\xi \ll N$ is always fulfilled and as such the overlap integrals couple a finite number of states which corresponds to an SRN and are characterized by glassy dynamics with nearly frozen local observables. If SRN features indeed are detected in disordered nonlinear systems then the glassy behavior could shed some light on the many-body localization. Thus far the Lyapunov spectra analysis shows a quick decay of the spectra in the case of $\xi \ll N$ which is a sign of SRN case. However we call for a more comprehensive study since the features of the Lyapunov spectra do not completely follow the SRN behavior observed before, such as the exponential decay of the spectrum.

### 2.4.3 Lyapunov spectrum scaling for Hamiltonian dynamics

So far we used the unitary map evolution as a toolbox for our studies due to the advantages they provide from the numerical standpoint. Realistic physical setups are described by continuous-time Hamiltonian dynamics. The numerical evolution of such systems is more costly and requires implementation of sophisticated algorithms to ensure the validity of conservation laws. In what follows we report on the application of the Lyapunov spectrum scaling analysis to a Hamiltonian setup with the system size of $N = 100$ sites.

In the Sec. 1.4.2 we discussed the networks of observables using the example of a chain of coupled anharmonic oscillators also known as the Klein-Gordon model:

$$H_{\text{KG}} = \sum_n \left[ \frac{1}{2} p_n^2 + \frac{\varepsilon}{2}(q_{n+1} - q_n)^2 + \frac{1}{2} q_n^2 + \frac{1}{4} q_n^4 \right]. \tag{2.48}$$

The proximity to integrable limit is controlled by the ratio of the energy density to the coupling constant $h/\varepsilon$. In the limit of weakly coupled anharmonic oscillators $\varepsilon \ll 1$, $h = const$ the system is in the SRN regime with the energy of each oscillator conserved. On the other hand in the case of low energy density $\varepsilon = const$ $h \ll 1$ we have an LRN of intercoupled normal modes.

We use the $ABA864$ symplectic integration scheme [111] and evolve the system of $N = 100$ sites up to times $T = 10^8$ with the relative energy error $E_r = (E(T) - E(0))/E(0) \approx 10^{-6}$. In Fig. 2.20 and Fig. 2.21 we plot the rescaled Lyapunov spectra for the system in SRN regime. The energy density is fixed $h = 1$ and the coupling assumes values $\varepsilon = \{0.1, 0.075, 0.05, 0.025, 0.01, 0.0075, 0.005\}$ from top to bottom. We notice the behavior similar to that displayed by unitary circuit maps in Fig. 2.14 and Fig. 2.15. As the system is tuned closer to the integrable limit the spectrum shows an exponential decay. In the Fig. 2.22 we fix the hopping term at $\varepsilon = 1.0$ and vary the energy density in the range $h = \{0.1, 0.075, 0.05, 0.025, 0.01, 0.0075, 0.005\}$, this is LRN regime. As opposed to the SRN limit there is no fast exponential decay, the result shows the features similar to those of an LRN case in unitary circuit map, see Fig. 2.16.

The results agree with the expectations set by our studies of unitary maps and provide a convincing argument towards the universality of the Lyapunov spec-tra scaling for weakly nonintegrable systems. The studies could be extended towards other integrable systems such as Toda chain, where the modes are extended but differ from those found in the linear system.

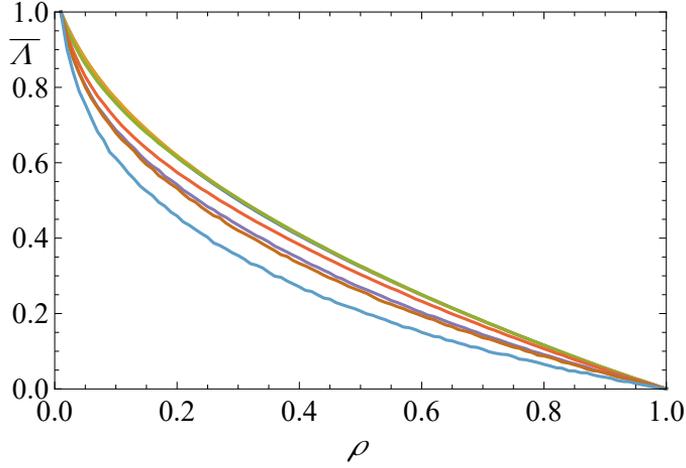

Figure 2.22: Renormalized Lyapunov spectrum $\{\bar{\Lambda}_i\}$ for LRN in proximity to the integrable limit. The hopping parameter $\varepsilon = 1.0$ is fixed. The deviation from integrable limit $h$ varies from $10^{-1}$ (top) to $5 \cdot 10^{-3}$ (bottom). For all cases the system size $N = 100$.

# Chapter 3

# Final remarks

The main achievement of the framework presented in this thesis lies in revealing the existence of classes of weakly nonintegrable dynamics. In recent years the interest to extremely slowly (or even non-) thermalizing dynamics has been growing with the discovery of quantum effects such as many-body localization. Understanding such dynamics is a challenging task for modern day physics. Some extensions to the classical disordered nonlinear systems have been made to reveal extremely slow glassy dynamics. It turns out that even in the classical realm there is more to uncover regarding the thermalization properties of weakly nonintegrable systems with large numbers of degrees of freedom. In this thesis we have formulated and presented in detail a novel framework for classifying such systems.

In particular the models with weak nonlinearity such as FPUT belong to a broad class of long-range networks. The nonlinearity serves as a perturbation to the dynamics of extended normal modes of a linear system and induces a coupling network with number of connections scaling with the system size. The chaotic dynamics fully characterizes the thermalization properties as the relevant ergodization time can be fully determined from the Lyapunov time $T_E \sim T_\Lambda$. In LRNs there is no characteristic distance between observables. This is well captured by the scaling of the participation number of the tangent vector with the system size $PN \sim N$. This finding indicates the participation of the whole network in the equilibration process.

The other studied class is that of short-range networks. This class is prevalent in the lattice systems with vanishing coupling constant. The coupling of observables has finite range which results in an inherent lengthscale emerging together with the timescale characteristic for chaotic motion. In proximity to the integrable limit most of the network is frozen apart from rare resonant spots where the dynamics is chaotic. Thus the shortest chaoticity timescale is much smaller than the timescale of ergodization of the whole system $T_\Lambda \ll T_E$. The thermalization takes place by means of diffusion of the resonances on characteristic distance. At variance to LRNs the tangent vectors are localized with constant participation numbers upon varying system size $PN = const$, signifying the presence of a lengthscale controlled by proximity to the integrable limit.

Some qualitative and quantitative differences between SR and LR networks were identified in the recent studies [45, 62, 49, 50, 51]. However the in-depth analysis of the influence of the choice of observables on the observed thermalization dynamics has not been performed. We find that the wrong choice of observables, such as those not matching to the actions in the integrable limit, might show quick thermal-like behavior which is captured by the decay of the width of the finite time averages distribution. Such behavior persists even in the limit itself when the actions are frozen. Without prior knowledge of the type of given observables, conclusions regarding the network type as well as the proximity to the integrable limit cannot be made [90].

We work around such ambiguity by proposing a framework of characterization of weakly nonintegrable dynamics by computing the full Lyapunov spectrum and its scaling properties as the system is tuned closer to the integrable limit [99]. Such analysis is the observable choice independent. Moreover the knowledge of the analytic form of actions of the integrable limit is not required to obtain reliable answers using this approach. The lengthscale free property of LRNs shows in the spectrum tending towards a limiting curve parametrized by a single relevant timescale – the largest Lyapunov exponent. In the SRN case, on the other hand, there exists an exponential decay of the spectrum signifying of additional lengthscale.

The analysis presented in this thesis applies to a range of setups including weakly nonlinear dynamics as well as those of vanishing coupling constants. However, broader research for the identification of new weakly nonintegrability classes is called for. We studied the dynamics of disordered nonlinear systems as well as free-propagating systems and obtained results requiring future investigation. Such setups together with cases of higher dimension, exponentially or polynomially decaying couplings, systems with nontrivial actions such as the Toda model point towards the future direction of research.